\documentclass[manuscript, screen]{acmart}
\AtBeginDocument{%
  }

\setcopyright{acmlicensed}
\copyrightyear{2018}
\acmYear{2018}
\acmDOI{XXXXXXX.XXXXXXX}
\acmConference[Conference acronym 'XX]{Make sure to enter the correct
  conference title from your rights confirmation email}{June 03--05,
  2018}{Woodstock, NY}
\acmISBN{978-1-4503-XXXX-X/2018/06}

\usepackage{multirow}
\usepackage[nice]{nicefrac}
\usepackage{algorithm}
\usepackage[noend]{algpseudocode}
\DeclareCaptionLabelFormat{noname}{#2}
\usepackage{subcaption}
\usepackage{fancyvrb}
\usepackage{fvextra} 

\newcommand{\MM}[1]{\textcolor{teal}{#1}}

\begin{document}


\title{The Emerging Use of GenAI for UX Research in Software Development: Challenges and Opportunities}

%

\author{Heloisa Candello}
\email{heloisa.candello@ibm.com}
\affiliation{%
  \institution{IBM Research}
  \city{S\~ao Paulo}
  \country{Brazil}
}

\author{Werner Geyer}
\email{werner.geyer@ibm.com}
\affiliation{%
  \institution{IBM Research}
  \city{Cambridge}
  \state{Massachusetts}
  \country{United States}
}

\author{Siya Kunde}
\email{skunde@ibm.com}
\affiliation{%
  \institution{IBM Research}
  \city{Yorktown Heights}
  \state{New York}
  \country{United States}
}

\author{Michael Muller}
\email{michael.muller@ibm.com}
\affiliation{%
  \institution{IBM Research}
  \city{Cambridge}
  \state{Massachusetts}
  \country{United States}
}

\author{Daita Sarkar}
\email{daita.sarkar@ibm.com}
\affiliation{%
  \institution{watsonx.governance, IBM}
  \city{Kochi}
  \state{Kerala}
  \country{India}
}

\author{Jessica He}
\email{jessica.he@ibm.com}
\affiliation{%
  \institution{IBM Research}
  \city{Seattle}
  \state{Washington}
  \country{United States}
}

\author{Mariela Claudia Lanza}
\email{mlanza@ar.ibm.com}
\affiliation{%
  \institution{IBM Research}
  \city{Tandil}
  \state{Buenos Aires}
  \country{Argentina}
}

\author{Carlos Rosemberg}
\email{carlos.rosemberg@ibm.com}
\affiliation{%
  \institution{AI Design, IBM}
  \city{Markham}
  \state{Ontario}
  \country{Canada}
}

\author{Gord Davison}
\email{gord.davison@ibm.com}
\affiliation{%
  \institution{AI for Product Teams, IBM}
  \city{Toronto}
  \state{Ontario}
  \country{Canada}
}

\author{Lisa Pelletier}
\email{lisa.pelletier@ibm.com}
\affiliation{%
  \institution{AI for Product Teams, IBM}
  \city{Austin}
  \state{Texas}
  \country{United States}
}



\begin{abstract}

The growing adoption of generative AI (GenAI) is reshaping how user experience (UX) research teams conduct qualitative research in software development, creating opportunities to streamline the production of qualitative insights. This paper presents findings from two user studies examining how current practices are challenged by GenAI and offering design implications for future AI assistance. Semi-structured interviews with 21 UX researchers, product managers, and designers reveal challenges of aligning AI capabilities with the interpretive, collaborative nature of qualitative research and tensions between roles. UX researchers expressed limited trust in AI-generated results, while product managers often overestimated AI capabilities, amplifying organizational pressures to accelerate research within agile workflows. In a second study, we validated an AI analysis approach more closely aligned with human analysis processes to address trust issues bottoms-up. We outline interaction patterns and design guidelines for responsibly integrating AI into software development cycles.
\end{abstract}



\received{20 February 2007}
\received[revised]{12 March 2009}
\received[accepted]{5 June 2009}

\begin{teaserfigure}
\centering
  \includegraphics[width=\textwidth]{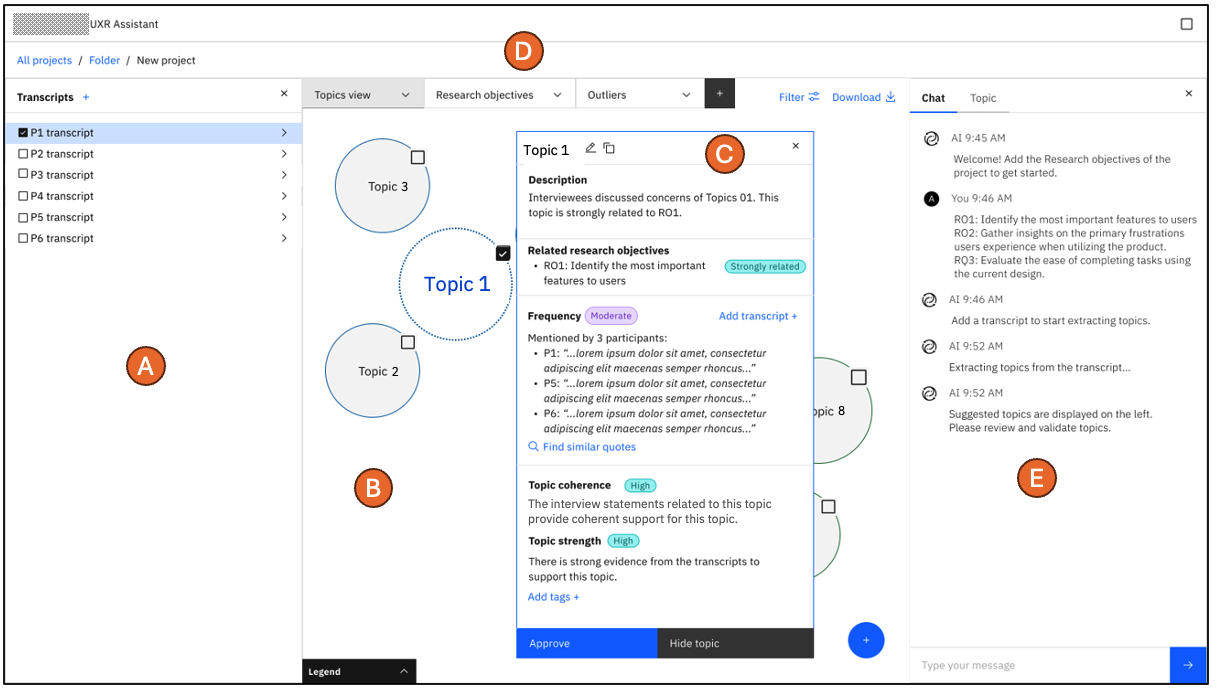}
  \caption{One of the five screens of the design probe we showed during semi-structured interviews with 21 UX researchers, product managers, and designers. The transcript area (A) shows interview data uploaded into the system and allows users to drill down into details (not shown here). The default topic view (B) renders topics extracted by AI as bubbles on a canvas. Clicking on topics displays a topic summary and details about a topic including metrics about the topic. (C). Users can switch between multiple views of the data (D) including an outlier view and a view organized by research topics. The chat interface (E) allows users to interact with the application and its data conversationally.}
  \Description{This image displays a user interface mock-up used during our design probe in an interview study with 23 UX researchers, product managers, and designers. It highlights five areas (A) through (E). The transcript area (A) shows interview data uploaded into the system and allows users to drill down into details (not shown here). The default topic view renders topics extracted by AI as bubbles on a canvas. Clicking on topics displays a topic summary and details about a topic including metrics about the topic (C). Users can switch between multiple views of the data (D) including an outlier view and a view organized by research topics. The chat interface (E) allows users to interact with the application and its data conversationally.}
  \label{fig:teaser}
\end{teaserfigure}
\maketitle

\section{Introduction}
Generative AI (GenAI) is being rapidly adopted across industries, transforming how professionals approach knowledge work and decision-making tasks \cite{geyer2025case, brachman2025current, lee2025impact}. In the context of agile software development, user research, in particular qualitative  analysis, plays an important role in translating customer needs and requirements into actionable insights for product teams \cite{Blandford2016, Khan2025}. This process is inherently interpretive and often collaborative, requiring iterative coding, synthesis, and also a deep understanding of user context \citep{braun2022conceptual, braun2023toward, bowman2023using}. For a long time, researchers have explored how technology can support qualitative analysis, with the goal to reduce its labor-intensive nature while preserving interpretive depth \cite{Blandford2016, Zimmerman2007, Kuutti2014, Blei2003, gaver2012should, Pietsch2018, Amershi2014, Paul2011}.

With the increasing accessibility of large language models (LLMs), user experience researchers (UXRs) have begun experimenting with incorporating GenAI into their workflows and practices. At the same time, the availability of GenAI creates mounting pressures to accelerate user research and align the analysis process with agile timelines and expectations---a longstanding challenge that pre-dates LLMs \cite{Kuang_2022, Jiang2021, Gray2018, Kuusinen2014, Larusdottir2012, BRHEL2015163}.
While the range of possible user research tasks that LLMs can assist with is wide, conducting thematic analysis as part of their practices is often considered one of the biggest challenges \cite{reyes2025augmented}. This includes using GenAI to assist with coding, identifying patterns, and synthesizing insights \citep{zhang2023redefining, Blei2003}. 

Recent academic work has begun to explore these forms of assistance. For instance,  \citet{naeem2025thematic} outline a step-by-step approach for integrating ChatGPT into thematic analysis, emphasizing enhanced transparency, speed, and the potential to reduce bias in qualitative coding. Other studies suggest that LLMs may be able to generate insights comparable to human coders and potentially reduce time and effort \citep{bryda2024words,de2025reflections, naeem2025thematic}.
Research has also revealed a growing but uneven adoption of GenAI in UX practice. For instance, \citet{de2025ideation} highlight how UX professionals experiment with GenAI in ideation and synthesis, often without formal guidelines or shared practices. \citet{heigl2025generative}'s case study shows the use of GenAI to support early-stage design work and rising concerns about human judgment and expertise. From a design perspective, integrating GenAI into UX workflows also introduces new interaction patterns \citep{wiberg2023automation}, but many questions remain about how these systems should be integrated into their day-to-day work, support trust and rigor, and adapt to the diverse needs of UX roles.

Despite emerging insights, there is limited empirical understanding of how different stakeholders in the UX research process envision GenAI in their everyday practice. This paper addresses that gap by investigating the current and emerging use of GenAI in UX research workflows in a large global company (anonymous during review), with a particular focus on AI-assisted qualitative analysis of interview data. Drawing on data from 21 semi-structured interviews with a design probe and a quantitative evaluation of an LLM-based topic extraction approach, our work maps current realities and challenges of GenAI use to possible future system designs through the lens of the following research questions:

\begin{enumerate}
    \item[RQ1:] How do the key stakeholders approach qualitative analysis for product design, evaluation, and communication to external stakeholders? (Study 1)
    \item[RQ2:] What are current uses, challenges, and opportunities for GenAI in qualitative user research for product development? (Study 1)
    \item[RQ3:] What user experience and interaction patterns do UX researchers envision for an effective GenAI integration into their workflow? (Study 1)
    \item[RQ4:] How feasible is a bottom-up LLM-based approach in performing thematic analysis similar to traditional human coding methods? (Study 2)
\end{enumerate}

This paper makes the following contributions:
\begin{itemize}
    \item We present results from an interview study (Study 1) that provide insights from two major stakeholder groups, UX researchers and product managers, about current research practices in a large global technology company, outlining challenges and opportunities for GenAI integration into user research.
    \item We offer design implications for future AI assistance from a design probe (Study 1), highlighting actionable directions for addressing key challenges in systems that support qualitative analysis.
    \item We present quantitative findings that validate a bottom-up topic extraction approach for thematic analysis, supporting traceable, human-in-the-loop system designs (Study 2).
\end{itemize}

While both UX researchers and product managers saw potential for GenAI to facilitate research, our interview results reveal key disparities between their approaches and views on thematic analysis, which created diverging needs for GenAI integration. UX researchers value interpretive depth and accountability for research findings. As such, they emphasized the need for mechanisms to support their agency, including traceability of AI-extracted insights, explanations, and features for human validation. In contrast, product managers prioritize speed and digestibility of insights, and hence were more optimistic and less critical about the use of GenAI to streamline research and present high-level findings. Our system evaluation of a bottom-up approach additionally reinforced improved outcomes by using a human-in-the-loop approach. Together, our findings indicate the need for role-sensitive GenAI support that preserves the rigor and control required by UX researchers while allowing results to be accessible to product managers -- especially in product development practices where diverse stakeholders collaborate closely to leverage research insights.

\section{Related Work}

\subsection{User Experience Research in Software Development}

User experience (UX) research plays a critical role in software product development by uncovering user needs, expectations, pain points, and opportunities that inform product design and direction. Qualitative data analysis (QDA), particularly thematic analysis, is an important component of this work, helping translate user narratives into actionable insights for software development teams.

UX research encompasses a diverse range of methods, each with distinct roles in software product development. \textit{Generative methods}, such as ethnographic fieldwork, contextual inquiries, diary studies, and 
semi-structured interviews, are applied early in the lifecycle to uncover user needs, behaviors, and contexts \citep{Blandford2016}. \textit{Evaluative methods}, such as usability testing, heuristic evaluations, and concept tests, assess interaction quality and whether user needs are met during iterative design phases \citep{righi2007ucdstories}. \textit{Longitudinal approaches}, including field deployments and retrospective assessments, capture how user experience evolves over time, particularly in software and interactive systems \citep{Karapanos2009,Karapanos2010,Roto2011}. 
Finally, \textit{participatory and co-design methods} integrate users directly into solution development, bridging research and design \citep{muller2012participatory}.

Despite its importance, UX research faces persistent challenges in contemporary product development. These challenges include reactive UX practices~\cite{Kuusinen2014}, prioritization of functional requirements over usability~\cite{BRHEL2015163}, and organizational buy-in~\cite{Goodman2012, Gray2018, Vaananen2008}. A central difficulty lies in aligning research timelines with agile and lean processes. For example, \citet[p. 1]{Larusdottir2012} found that in scrum projects, short sprint cycles and emphasis on incremental delivery left little time for involving users or conducting UX evaluation, leading to situations where "the big picture of UX is often lacking".
In qualitative analysis specifically, scaling represents a persistent obstacle: \citet{Jiang2021} emphasize that qualitative interpretation remains inherently messy and resource-intensive, even when supported by computational tools, while \citet{Kuang_2022} show that UX practitioners struggle under time pressure to collaboratively synthesize large volumes of data into actionable insights. More broadly, translating unstructured data—such as interviews and field notes—into product decisions remains a bottleneck across software contexts. Finally, organizational buy-in is often challenging: unless UX research outputs are framed in ways that resonate with business priorities and decision-makers, they risk being sidelined in favor of short-term delivery pressures. Prior work has shown that UX findings are frequently overruled when organizational goals emphasize growth or speed over user needs \citep{Gray2018}, and practitioners note that research insights must be packaged in stakeholder-relevant, persuasive ways to gain traction \citep{Goodman2012}. Moreover, even when organizations value user experience in principle, evaluation methods must be pragmatic, resource-sensitive, and produce results meaningful to multiple stakeholders in order to be taken up in practice \citep{Vaananen2008}.

Proposed solutions have largely focused on strengthening collaboration, methodological rigor, and communication, with additional efforts aimed at integration into agile workflows and, more recently, computational augmentation. Collaborative approaches such as team-based thematic analysis and the use of shared analysis environments distribute analytical tasks while helping maintain reliability \citep{Blandford2016}. 
Well-known
frameworks, most prominently Braun and Clarke’s diverse practices
of thematic analysis, may provide methodological consistency when research is conducted by non-specialist roles such as product managers and designers 
\cite{braun2022conceptual, braun2023toward}.
Storytelling and visualization strategies—such as journey maps, affinity diagrams, and other synthesis artifacts—have been shown to enhance organizational buy-in by making qualitative insights more tangible and actionable for diverse stakeholders \citep{Kolko2010}. Beyond these methodological supports, researchers have proposed ways of aligning UX practices with agile and lean development, including “design one sprint ahead” and embedding UX specialists directly within scrum teams \citep{Kuusinen2014,BRHEL2015163}, an approach that does not always consider the sensitivity and contextualization that UX researchers want to preserve. Finally, emerging work explores the role of computational augmentation, where AI and LLMs support the coding, clustering, and synthesis of qualitative data, offering potential pathways for scalability while raising new questions of rigor and validity \citep{goyanes2025thematic}. 


\subsection{AI-Assisted UX Research}

The integration of artificial intelligence (AI) into UX research has followed an evolutionary trajectory from early explorations of interactive machine learning to today’s use of LLMs. \citet{Zimmerman2007} and \citet{Kuutti2014} laid early conceptual foundations, emphasizing practice-based inquiry and research-through-design as approaches for situating computational tools within qualitative and design research practices. Building on these ideas, \citet{Amershi2014} introduced the paradigm of interactive machine learning, where humans iteratively provide feedback to shape and refine models. In UX research, this approach foregrounded questions of transparency, control, and usability in adaptive systems \MM{\cite{jameson2007adaptive}}, and also offered researchers a methodological tool for conducting human-in-the-loop qualitative analysis.

Before the recent rise of LLMs, researchers explored computational methods to assist thematic analysis and coding of qualitative data. Early work on topic modeling, particularly Latent Dirichlet Allocation (LDA) \citep{Blei2003}, showed how unsupervised learning could uncover latent themes in large text corpora, and subsequent studies adapted these approaches to social media and health domains \citep{Paul2011}. For user research contexts, Pietsch and Lessmann \citep{Pietsch2018} demonstrated that short-text topic models such as the Biterm Topic Model (BTM) and Word Network Topic Model (WNTM) could scale the analysis of open-ended survey responses, offering a viable—though imperfect—alternative to manual coding. Their study highlights both the promise and limitations of pre-LLM computational support: while such models improve topic coherence \citep{roder2015exploring} and document classification over LDA \citep{Blei2003}, they remain limited in producing fine-grained, one-to-one mappings of responses to themes, requiring human interpretation for labeling and synthesis. Collectively, these approaches foreshadow the role of AI as an analytic partner in thematic analysis, paving the way for more interactive and semantically rich systems in subsequent years.

Recent work has accelerated these trajectories by positioning LLMs as promising tools for augmenting qualitative and design practices. Studies show that LLMs can lower barriers for participation and design ideation: \citet{Stadler2025} demonstrate how they enable both experts and laypersons to engage in scenario writing, while \citet{Krajcovic2025} show that LLM-powered conversational agents improve participant engagement and data quality in surveys. LLMs also demonstrate potential for improved context awareness, resulting in computational support that is more sensitized to study data and researchers' preferences than preceding tools~\cite{ye2025scholarmate}.
Beyond participation, LLMs are being tested in applied workflows. For instance, \citet{Gomez2025} present a case study in content management systems where GenAI supported evaluative and assistive tasks, such as keyword analysis and content optimization. While domain-specific, it illustrates how AI tools can be embedded into everyday research and design workflows. 

At the same time, prior work also describes structured approaches to augment portions UX researchers' workflows while ensuring rigor and transparency. \citet{goyanes2025thematic} propose a six-phase protocol for conducting thematic analysis with ChatGPT, including a validation checklist, while \citet{Naeem2025} provide a detailed step-by-step process with prompt templates aligned to Braun and Clarke’s framework. Complementing these, \citet{Xiao2023} benchmark LLMs against traditional supervised and unsupervised NLP, showing fair-to-substantial agreement with human coders and highlighting challenges around prompting, auditability, and bias. More workflow-focused efforts include CollabCoder \citep{CollabCoder2024}, which integrates LLMs into collaborative open coding and codebook development while addressing risks of over-automation, and LLooM \citep{Lam2024}, which frames LLMs as lab-in-the-loop co-analysts capable of supporting ideation, summarization, and meta-analysis in empirical projects. 

Building on these prototypes, researchers are experimenting with multi-agent systems to scale analysis. Rasheed et al. \cite{rasheed2024largelanguagemodelsserve} propose a multi-agent architecture where specialized LLM agents automate different types of qualitative analysis, while Sankaranarayanan et al. \cite{Sankaranarayanan2025} design a multi-agent pipeline that mirrors human thematic analysis stages and argue for validity metrics and transparency. Applied evaluations also expose limitations: Khan et al. \cite{Khan2024}, in a pilot on the controversial Australian Robodebt case, found overlaps and divergences between GPT-4, LLaMA-2, and human coders, stressing that LLMs should augment, not replace analysts, particularly in sensitive contexts. Finally, prompt-engineering studies \citep{Wang2023} compare strategies for topic modeling with LLMs, demonstrating trade-offs in coherence, diversity, and reproducibility. 

Together, these studies outline a historical progression: from early topic modeling and interactive machine learning augmenting qualitative coding, to today’s LLMs and multi-agent systems serving as co-analysts, facilitators, and evaluators of UX research. While methodological and ethical concerns remain—particularly around transparency, reproducibility, bias, job loss, and deskilling 
—AI is increasingly positioned as a complement to traditional qualitative methods, enabling richer, more scalable, and context-sensitive UX research \citep{Xiao2023,goyanes2025thematic,CollabCoder2024}. Looking ahead, a central challenge is to establish rigorous validation practices, effective triangulation with human judgment, and integration into collaborative team workflows, ensuring that AI assistance strengthens rather than undermines the credibility and impact of UX researchers and their work.

In our project, we plan to build LLM-based tools to augment the work of staff who perform user research in industrial settings, and with their stakeholders. Therefore, beyond demonstrations of technical feasibility \cite{goyanes2025thematic, naeem2025thematic, Xiao2023}, we wanted to understand the work practices of these employees - both their current (or as-is) practices, and their expectations and desires for AI-supported future (or to-be) practices that they envision. To do that, we performed a qualitative study of the work of current user researchers and product managers (Study 1), and we tested one key algorithmic hypothesis to support future LLM-based systems (Study 2).

\subsection{Research Stakeholders}

While computational augmentation and AI assistance may yield scalability, the effectiveness of integrating technology in practice also depends on the organizational context and the diverse stakeholders who conduct, interpret, and act upon its findings. Typically, at the center of user research activities are \textit{UX Researchers} (UXRs), trained specialists who apply qualitative and mixed-method approaches to uncover user needs and translate them into actionable insights\cite{robinson2018past}. In larger organizations, they collaborate closely with product managers, designers, and engineers, whereas in smaller or resource-constrained settings, their responsibilities are often absorbed by other roles. The presence or absence of dedicated UX researchers strongly shapes how research is conducted and valued in software development: without them, research tasks are often distributed across product managers, designers, or developers, leading to fragmented practices and reduced methodological rigor \citep{Larusdottir2012, BRHEL2015163}. Industry reports similarly suggest that UX research is structurally under-resourced, with typical researcher-to-designer ratios around 1:5 \citep{Kaplan2020}. This dynamic underscores how organizational structures determine not only the visibility of user perspectives, but also the feasibility of adopting emerging practices such as AI-assisted analysis into mainstream UX workflows.

\textit{Product managers (PMs)} are primarily responsible for defining product vision, prioritizing features, and aligning development with business goals. Their role centers on managing product road-maps, balancing stakeholder needs, and ensuring delivery meets strategic objectives. In relation to user research, PMs are not typically trained specialists, but in contexts lacking dedicated researchers, they often act as facilitators or surrogates. They may conduct lightweight interviews, usability sessions, or feedback analysis. Industry evidence suggests this practice is widespread: Maze’s 2025 survey of 800 product professionals found that 42\% of product managers report conducting user research, often alongside designers (70\%) and UX researchers (63\%) \citep{Maze2025}. 
However, without formal research training and with their dual responsibility as business strategists, critics warn that they may compromise validity and introduce bias in their research~\cite{Levitt2022}. Practical accounts and academic work in software engineering document common forms of cognitive bias in product research (e.g., selection effects, confirmation bias, and procedural bias), which can distort findings when research is conducted by non-specialists \citep{Karapetyan2024, Mohanani2017}. 

 We briefly mention several other roles who interact with UX Researchers and Product Managers. We did not interview people in these roles. We include them for context. \textit{Designers'} responsibilities emphasize ideation, prototyping, visual communication, crafting consistent user experiences, and usability assessments 
\cite{Blandford2016,Nielsen2014}. \textit{Engineers'} day-to-day work centers on coding, debugging, and optimizing software to meet product requirements. In some organizations, Engineers may also be called Developers
\cite{Ko2006, 
Gunatilake2024}. Finally, \textit{business stakeholders} play a crucial role in determining whether user research insights are adopted. Empirical studies of UX integration show that active stakeholder involvement, particularly from managers and decision-makers, is a key success factor for embedding UX practices in software development \citep{Kashfi2016}. 

UX research is a cross-functional endeavor in which UX researchers provide methodological rigor, but its impact depends on the involvement of PMs, designers, engineers, and business leaders. Their shared priorities and practices ultimately determine whether and how research insights are integrated into product strategy or sidelined by short-term delivery pressures. 

Against this backdrop, we are exploring emerging AI-assisted approaches that can offer new possibilities for scaling analysis and embedding user perspectives more effectively across diverse stakeholder roles. The above stakeholder roles and dynamics are comparable to roles in our organization where we conducted our study. Our research focused only on PMs and UXRs (including designers who practice user research). We empirically provide deeper insights into the different challenges, pressures, and priorities they face in user research workflows and their current use of GenAI. Their differences in needs and expectations informed implications and suggestions for the design of future tools for AI-assistance for qualitative analysis.



\section{Overall Methodology}

We conducted two user studies to examine prevailing practices in qualitative analysis and the emerging use of GenAI, with the aim of deriving design implications for future AI-assistance. Our approach integrated qualitative inquiry (Study 1, RQ1--3) with a quantitative evaluation of bottoms-up AI assistance (Study 2, RQ4), allowing us to assess both current practices and future possibilities. In Study 1, we employed semi-structured interviews in combination with a design probe.  We recruited 23 participants engaged in professional software development, including UX researchers (UXRs), product managers (PMs), and designers from different business units within our company. The participants were selected to represent a diverse range of roles and levels of experiences with qualitative research. Study 2 explored the feasibility of a bottoms-up, LLM-driven approach to thematic analysis. This study was motivated by findings in the first study that pointed to diverse trust-related challenges with top-down LLM approaches. 
This study represents an initial feasibility assessment of an algorithmic foundation for a bottom-up approach.


\section{Study 1: Methodology}

\subsection{Procedure}
To address RQ1--3,
we conducted a series of semi-structured interviews including 
a visual design probe \cite{he2024ai}. The study sought to examine several key areas: (1) the roles of product team members in user research, (2) their end-to-end processes and practices for conducting qualitative studies, (3) the methodologies employed for data analysis, (4) the practices for collaboration and communication related to user research, (5) the current use and challenges of AI in user research, and (6) potential applications of artificial intelligence to enhance qualitative user research, with particular emphasis on thematic analysis (detailed interview guide in Appendix \ref{appendix:guide}).

The design probe was inspired by existing qualitative analysis tools and incorporated some envisioned AI features to hypothetically support qualitative analysis. Figure \ref{fig:teaser} shows one of the five screenshots we used during the interview with AI-extracted topics in the center and a detailed topic description. The remaining screenshots are in the Appendix---Figures \ref{fig:screenshot1},\ref{fig:screenshot3},\ref{fig:screenshot4},\ref{fig:screenshot5}. The probe was intended to elicit reflection, reactions, and critical opinions, rather than to present a comprehensive vision of an AI-assisted analysis tool.

All interviews were conducted by one moderator and at least one note-taker via video-conferencing and took between 60-75 minutes. Sessions were recorded, and participants provided informed consent, including agreement for using AI for analysis. Study participation was voluntary, with the option to withdraw anytime.Data was de-identified, stored securely, and analyzed anonymously. For the AI analysis (Section \ref{sec:system}), all data were processed using internal corporate LLM services; no data was used for model re-training. The study was conducted in accordance with corporate policies on internal user research and data handling. As compensation for their involvement, participants received a gift equivalent to USD 25.

The authors conducted thematic analysis \cite{Braun01012006, Braun2013} of the interview transcripts. After analyzing two interviews, three authors created an initial code book. The code book (Appendix - Table \ref{apendix:codebook_1}) was subsequently applied to all the transcripts
and incrementally extended if any new themes were identified, in
agreement with all analysts. Each transcript was analyzed by two different authors and codes were agreed upon in a discussion. 



\subsection{Participants}
We recruited 21 participants through internal company communication channels. 
(57.14\% female , 42.86\% male). 
The participants, all employed at the same global technology company, included 12 UX Researchers, 4 Product/UX Designers, and 5 Product Managers. They represented departments such as Software, Infrastructure, and Consulting, and were distributed across 6 countries. 
Participants had between 2 and 14 years of experience with UX research and thematic analysis (M = 6.27 years). Designers reported hands-on involvement in conducting research, i.e. acting as user researchers, while Product Managers typically collaborated with UXRs or conducted basic research as part of their role. Most participants focused on qualitative methods, involving both evaluative concept testing and generative  discovery research. Key collaborators included Product Managers, Designers, Developers, and UX Researchers; nearly half (9 of 21) also engaged with Sales and Marketing teams.

\begin{table*}[h]
  \caption{Participant Details}
  \label{tab:participants}
  \begin{tabular}{llll}
  \toprule
    \textbf{Participant No.} & \textbf{Role} & \textbf{Country} & \textbf{Organisation} \\
    \midrule
    P1 & Senior UX Researcher & United States & Software \\
    P2 & Lead UX Researcher & Ireland & Software \\
    P3 & UX Researcher & United States & Infrastructure \\
    P4 & UX Researcher & United States & Software \\
    P5 & Technical Product Manager & United Kingdom & Software \\
    P6 & UX Researcher & India & Software \\
    P7 & UX Researcher & United Kingdom & Software \\
    P8 & Lead UX Researcher & United States & Infrastructure \\
    P9 & Innovation Designer & Germany & Consulting \\
    P10 & GTM Product Manager & India & Software \\
    P11 & Product Manager & Ireland & Software \\
    P12 & Lead UX Researcher & United States & Software \\
    P13 & PLG Product Manager & India & Software \\
    P14 & Design Manager & Uruguay & Consulting \\
    P15 & UX Researcher & United States & Software \\
    P16 & UX Designer & Germany & Software \\
    P17 & Lead UX Researcher & Ireland & Software \\
    P18 & Product Manager & United States & Software \\
    P19 & Senior UX Designer & United States & Software \\
    P20 & Lead UX Researcher & United States & Software \\
    P21 & UX Researcher & United States & Software \\
    \bottomrule
    \Description{A table listing 21 participants involved in a UX-related study. The table has four columns: Participant Number, Role, Country, and Organization. Roles include various UX positions such as UX Researcher, Lead UX Researcher, Senior UX Designer, UX Designer, Design Manager, and Product Manager. Participants are from diverse countries including the United States, Ireland, United Kingdom, India, Germany, and Uruguay. The majority are affiliated with Software organizations, with a few from Infrastructure and Consulting sectors. The table is structured in rows, each representing one participant, and is used to illustrate the diversity of roles, geographic distribution, and organizational contexts within the study.}
  \end{tabular}
\end{table*}

\section{Study 1: Results}
We present the themes we identified for RQ1--RQ3 while RQ4 will be discussed in Section \ref{sec:study2results}. 

\label{sec:results-rq1}

\subsection{RQ1 - How do the key stakeholders approach qualitative analysis for product design, evaluation, and communication to external stakeholders?}

The two principal roles that we interviewed were User Experience Researchers (UXRs) and Product Managers (PMs). Because these two roles work closely together, we had anticipated that they would take similar approaches to how they worked with user data, and particularly with analytic approaches to interview data. We reasoned that they were using the ``same data,'' and that they would use those same data in the same ways. We were wrong.

\subsubsection{Working Relationships}

While people in both roles work in teams, the content of their work is of course different. We learned that their working relationships are also different: UXRs focus largely within teams, while PMs work with other teams, clients, and customers. We use the word ``\textbf{team}'' to indicate the formal organizational structure in which people work, and we will use the word ``\textbf{circle}'' to indicate the additional stakeholders that formed the working-context for each person's actions. P17 (Lead UXR) noted that PMs manage their circles by \textit{``own[ing] a lot of relationships with customers as well. Depending on which business unit you're in, you might have sales, CSMS, go to market depending on the product who would be involved. Development, of course, because it's all about, of course, feasibility you need to know.''}

In discussing these diverse working relationships, the most frequent roles mentioned by UXRs were "PMs, UXRs, researchers, managers, and developers'' - i.e., members of their teams. By contrast, the collaborative roles mentioned by PMs were a much flatter distribution, with roughly equal numbers of UXRs and members of Sales departments as their most frequent collaborators - i.e., their frequent collaborators came from both their own teams and their circles (beyond their teams).

\subsubsection{Meetings}

Despite these differences, both roles frequently engaged in what were formally called "3-in-a-box" team meetings (UX, PM, developer \citep{Samsonov2025ProductTriad}).
In some cases, more stakeholders were involved:
the size of each of a series a meeting might grow incrementally with greater knowledge: \textit{``Once that playback's done again, we come back to the main stakeholders... or the small team involved first, do a little playback, see if there's any tweaks they want... Wider playback and making our findings available to [company staff] more generally''} (P2, Lead UXR). 

UXRs often chose diverse formats for their meetings, depending on purpose and membership. P3 spoke of convening \textit{``a workshop. Not only with the PM as sometimes I invite other stakeholders... like the design lead maybe development [to] discuss OK what do we know about this user? What are [people] thinking about it? This problem? What are the assumptions? What are the hypotheses?''} For project ``convergence points,'' P9 (Lead UXR) said \textit{``I bring the three in the box in or any other stakeholders really...''} Flexibility appeared to be key. P1 (Senior UXR) said that, for 3-in-a-box meetings, \textit{``it depends a lot on the project, in the road map ahead. If it's a tight schedule, then everyone's invited, right?''} 

By contrast, only two PMs mentioned 3-in-a-box meetings as collaborative practices. PMs' meeting practices were more likely to be shaped by the expectations of their wider circles. For example, P5 (PM) described types of larger meetings, such as discussions beginning with \textit{``this is the requirement,''} and a subsequent meeting in which \textit{``our researcher would come back with the set of questions in terms of, are these making sense? I did those things and so on. So that is second level.''} A follow-on meeting could involve a \textit{``third stage where we kind of finalize the use the journey and the Figma click-through prototype and then share it with the user research team.''} As with UXRs, for PMs, flexibility was also key. However, their flexibility involved different kinds of meetings in their broader circles.

\subsubsection{Working Practices}

Debriefs were a particular type of meeting that was important for UXRs, in part because they  \textit{``get[] people's... creative juices flowing and... get[] them to think about the problem in different ways''} (P2, Lead UXR). However, UXRs were aware that their principal ``client'' in those meetings - the PMs - had very little time. P18 (PM) remarked, \textit{``I only join the interviews when I have time and I think increasingly, unfortunately I'm finding it's hard for me to attend,''} And yet, debriefs were a way to engage PMs, so that they would become \textit{``really interested''} (P6, UXR) or \textit{``excited''} (P7, UXR) in the rich information from user research. UXRs limited the time or the frequency of debrief meetings as a strategy to accommodate PMs' schedules. 

As above, UXRs tried to be flexible not only with scheduling debriefs, but also in the ways that they conducted them. P18 (Lead UXR) reported using \textit{``a slide deck presentation share out. I have an appendix with... more...'}' in a formal meeting with 50\% presentation time. P14 (Design Manager) reported using a spreadsheet as a shared medium to ask \textit{``what patterns did they notice?''} Other UXRs made the debrief more engaging and collaborative. P12 (Lead UXR) used \textit{``a Mural [to] make sure that we get everyone's stuff,''} and P14 (Design Manager) used the same product \textit{``to help everybody to do their insights''} in a staged sequence of actions, moving in a staged sequence from private annotation to group discussion, and \textit{``then you get again patterns with all together.''} P4 (UXR) added \textit{``I am a fan of the manual method of just doing it in Mural... You can see the notes, you can see when things were grouped. And then you can... source it back to the participant from set interview in the data logger and you can then pull in the interviews.'' }

\subsubsection{Analysis Activities}
\label{sec:analysis-activities}

In our company's engineering culture, employees are deeply committed to analysis as a basis for decisions. Both PMs and UXRs used analytic practices. People in each role used practices that suited their discipline and their timeframe.

Some UXRs preferred to do thematic analysis as a solitary activity, to be shared as a completed deliverable. P7 (UXR) described such a practice as \textit{``So basically all the the questions... I would replicate these on my Excel board... I created a template for me to to work... faster [so that I] don't have to create this every time.''} P12 (Lead UXR) also used solitary tagging of excerpts from interviews: \textit{``if it's a project that's relatively straightforward, it can be a lot easier because you're you're really just grouping... So it's a pretty straightforward tagging process.''} The solitary analysts often used structured methods to organize their observations, such as Airtable.

Others, like P6 (UXR), saw value to a team-participatory approach to thematic analysis: \textit{``everyone has a different point of view and everyone has something to add to these things. So we try doing that as much as possible.''} The medium that each UXR chose for the analysis reflects these diverse practices. As we described above, P12 (Lead UXR) and P14 (Design Manager) used Mural as a shared online graphical board for group mark-up via sticky-notes, transforming collaborative thematic analysis into collaborative annotation. The more solitary practitioners used Excel and Airtable, tending toward more formal analytic documents.

UXRs' descriptions of thematic analysis varied widely. A few referred to predetermined categories, such as \textit{``I kind of have my themes already developed. So basically buckets where I was going to put all those different themes and then within those buckets then I created specific topics so I knew what I was going to be asking''} (P1, Senior UXR). Others spoke of actively clustering quotations and observations without a preplanned schema. We repeat a quotation from P7 (UXR), with a different emphasis: \textit{``we want to discover what's there and we are going there pretending that we do not know anything about it. Then maybe I would do a very classic thematic analysis, so gathering, clustering and identifying trends.''}\footnote{In their recent work, \citet{braun2022conceptual, braun2023toward} carefully included preplanned categories including codebook-based analysis, as well as emergent and constructed themes.} Again, we observe the theme of UXRs' flexibility - suiting the analysis to the situation, the collaborators, and the preferences and strengths of each skilled practitioner.

As we wrote above, we had anticipated that UXRs and PMs would share a basis in practice, including similar methods in thematic analysis. However, only two PMs mentioned thematic analysis, and only in response to our questions. P5 (PM) said, ``the term itself is new to me,'' and P18 (PM) acknowledged that ``personally I find that the term thematic analysis [is] confusing.'' Those same two PMs and their PM colleague P13 hoped someday to use large language models to produce a summarization of interview contents.

We think it is fair, in this case, to move from the word ```summarization'' to the word ``summation.'' UXRs were interested to develop practices that made use of AI tools under their control (see Section \ref{sec:rq2}), to augment their human-skilled thematic-analysis practices. They hoped to increase their organizational effectiveness through faster analyses - but with a concern for risks related to quality, nuance, empathy, and non-verbal information. PMs had different needs. Their positions generally required them to ask the UXRs to summarize their results into client-sized bites of information. In contrast to AI-supported thematic analysis, PMs seemed to want an AI to produce those bites of information. We say none of this as criticism. We report the pragmatic pressures on each role of skilled professionals, and their current and hoped-for accommodations to those pressures.

\subsubsection{Summary of RQ1}

We learned that UXRs and PMs face different challenges, pressures, and priorities in their work, even though they often work together on important projects. We learned that their collaboration spheres were in general different - the distinction between Teams and Circles - and we learned that aspects of their needs were derived by their distinct workplace identities. Thereby, their needs from a qualitative analysis tool turned out to be rather different. These differences informed our implications and conclusions in the Discussion section.

\subsection{RQ2: What are current uses, challenges, and opportunities for GenAI in qualitative user research for product development?}
\label{sec:rq2}

\subsubsection{Current AI usage}
Participants widely experimented with GenAI in their workflows, using tools, such as, for example, Co-Pilot, Airtable, WatsonX with some having familiarity with tools like Dovetail or Maze.
 
Reported use cases spanned planning and preparation (e.g., discussion guides, research plans), domain/market scans, transcript summarization and cleanup, theme/topic/code/tag extraction, persona drafts, and presentation and communication support.
For example, P4 (UXR) creates surveys with AI because it's \textit{`` manual work and it takes so much time [..] the demographics,[..] you already have this information, but what you don't have is [..] questions to ask the participant, like how many years of experience?''}


For analysis, participants experimented with both top-down (requesting themes/topics and evidence from one or more transcripts, e.g. \cite{goyanes2025thematic}) and bottom-up (generating quote-level codes and grouping them into themes) strategies for thematic analysis. 
P12 (Lead UXR) used top-down outputs in a slightly different way as a tentative scaffold to make \textit{``it faster for me at that moment''} and then refine and interpret by \textit{``[going] into the interviews again and find all of [the] quotes or you know data that fits to these clusters.''}

\subsubsection{Challenges of AI usage}
While participants acknowledged GenAI's potential, experiences were mixed. Positive uses clustered around low-interpretation tasks, for example, polishing emails or reports, structuring and transforming data, and light domain scans.

Many described using AI cautiously, trying to have a human, unbiased opinion before considering AI results and validating outputs extensively by returning to the raw data. Some felt outputs were not always representative of the verbatim transcripts, tended to surface obvious rather than meaningful information that lacked alignment  relevant goals of the analysis. Participants often  perceived AI models as missing nuances, not capturing the real meaning and sentiment of user's words, and not being able to do a deep interpretative analysis, reinforcing the need for human validation. As P18 (PM) put it, \textit{``you lose some of the specific Nuggets when you summarize all the transcripts [..] AI [shouldn't] be used to derive insights without also having the humans [..] involved [..] because it doesn't have that same level of understanding of all the pieces.''}

The majority of participants reported low trust in AI for insight generation  due to the lack of traceability, inconsistent  outputs, missing content, and hallucinated, inaccurately attributed user quotes. P09 (Innovation Designer ) noted, \textit{``sometimes it gave me back quotes that were [..] were just wrong.''} and P1 (UXR) questioned net value:  \textit{``[It] missed important points to put into the summary [..] I'm kind of struggling in that whole is it really saving me time''}.

%

Several participants also raised ethical risks around data security, privacy, and bias due the lack of representative data of foreign languages in AI models. Participants also emphasized the importance of domain knowledge to properly interpret and evaluate AI outputs. Without it, AI can sound persuasive but be hollow. P17 (UXR) \textit{``if I didn't know the product [..] I'd be like, sounds great [..] but it's just fluff.''}

Some participants also found the process of both crafting effective prompts and manually organizing context information (e.g. transcripts)time-consuming and cognitively demanding. For experienced researchers, the effort of structuring AI inputs led to deep familiarity with the data - making  manual analysis more efficient. In other cases, limited conversational memory hindered iterative analysis. P09 (Innovation designer): \textit{`` it fails in remembering context [..] So you have to like, write the entire prompt again.''}


\subsubsection{Envisioning the future of AI usage}
Participants expressed both promise and concern about how GenAI may reshape user research. While ethical, methodological, and organizational challenges continue to be an important concern for participants, many 
envision AI as a capable partner in qualitative analysis offering new ways to accelerate research workflows without compromising quality.


Recurring concerns for future core UX practices included lack of contextual knowledge, nuance, and empathy. As P09 (Innovation designer) noted, \textit{``It can miss patterns [..] because people use different words or might miss something that looks like an outline, comparing to external context-sensitive info.''} Others cautioned against over-reliance: once AI produces an output, stakeholders may treat the job as “done.” P2 (UXR) remarked, \textit{``People are cognitive misers [..] following up an LLM feels like [the work is already done]  I don't think there's any real solution to getting people to do that [manual check].''}

Participants also flagged organizational and ethical risks: job displacement, low-quality analysis by untrained stakeholders, over-confidence in AI-generated results, reduced human validation, and deficits in transparency, privacy, and guardrails. Despite this, many expected AI to become a permanent part of the workflow—driven by both pressures to streamline research and the believe that trust could grow with stronger human-in-the-loop practices, multilingual models, and better explanations and interpretability.






Concrete ideas for future features focused on pairing speed with verifiability:   AI that surfaces relevant data with linked raw evidence for easy validation, a searchable repository tying research questions to quotes to aid in in structuring/grouping snippets to showcase as a video during playbacks/debriefs and reuse of past studies to detect historical patterns.
Crafting prompts was seen as productivity-critical—\textit{“make or break”} (P9, Innovation designer)—spurring suggestions for a prompt library to help novices (see \cite{wang2024promptlibrary}) and a tutorial-style assistant that, for instance, flags leading questions during live interviews. Participants also envisioned unifying persona development across teams, and PMs were eager to extend AI beyond study analysis to scan public data for patterns or even simulate users. As P5 (PM) put it, \textit{`` I wish [..] I could ask AI to [..] act as a user [..] point out something about this [..] that could [..] accelerate at least PM's product managers thinking and assessment of the problem''} pointing to a broader, more explorative role of AI in user research.

\subsubsection{Summary of RQ2}
Both PMs and UXRs recognized the potential of GenAI in qualitative research but perspectives often diverged. PMs were generally more optimistic, focusing on opportunities to accelerate workflows, whereas UXRs approached AI more critically with greater caution, emphasizing the importance of interpretive nuances, domain knowledge, and human oversight. 

\subsection{RQ3 - What user experience and interaction patterns do UX researchers and product managers envision for an effective GenAI integration into their workflow?}

Participants explored a visual probe for AI-assisted qualitative analysis, providing detailed feedback on the features and capabilities displayed in  Figures \ref{fig:teaser}, \ref{fig:screenshot1}, , \ref{fig:screenshot3}, \ref{fig:screenshot4}, \ref{fig:screenshot5}. 
Feedback summarized in the following themes revealed nuanced expectations regarding traceability, explainability, interaction modes, and control, as well as preferences for views aligned with individual sense-making styles--often mirroring the challenges surfaced in RQ2. 

\subsubsection{Traceability and Evidence as Foundations for Trust}
Participants repeatedly asked to anchor high-level insights in raw data via quote highlighting, transcript annotations, and bidirectional links.
As P9 (Innovation Designer) put it, they wanted the transcript \textit{``the parts in the transcript that kind of went into that topic [to be] highlighted''}, and P18 (PM) sought a way for \textit{"tying [] things [back] to my questions."} UXRs consistently pushed for grounded, inspectable connections between themes and evidence; some PMs were satisfied with concise summaries and fewer topics. P6 (PM) noted the probe was \textit{`` doing fairly good job [..] it does 90\%''} but preferred \textit{``top three''} topics over longer lists; P11 (PM) echoed a quality-over-quantity stance, hoping that \textit{``limiting the number of topics will increase the quality.''}

Evidence mattered at handoff: both PMs and UXRs wanted tooling that makes presentation effortless (e.g., quotes with links or clips). Participants asked for shareable, read-only views: P5 (PM) imagined copying a view link so stakeholders \textit{``landed [on] the same view,''} P6 (UXR) would \textit{``just put up this screenshot''}, and P19 (Senior UX Designer) said a read-only share \textit{``would replace a presentation.''} Several expected PMs to consume results rather than co-analyze. As P12 (Lead UXR) reflected: rather than \textit{``diving into [the] program,''} many PMs would want the analysis rendered \textit{``to a deck.''}

\subsubsection{Need for Explainability of AI Reasoning and Metrics}
Participants repeatedly asked why the system produced a given output--why a quote belonged to a topic, why a score was high or low, and what terms like "coherence" actually meant. P12 (Lead UXR) needed \textit{``strong evidence''} because a \textit{``'topic isn't necessarily a conclusion''}, while P5 (PM) expected the AI to \textit{``explain [..] why [it] arrived at this conclusion''}.


While basic metrics like frequency were clear and valued, derived metrics like coherence, strength, and color coding often produced confusion or incorrect mental models. P12 (Lead UXR) said \textit{"I'm not sure what topic coherence would [..] necessarily mean? [..] I would also be expecting that it wouldn't [..] show a topic if there wasn't high coherence [..] how I would use that?} Discoverability also mattered: P11 (PM) noted \textit{``that [..] coherence and strength [..] might be a bit confusing without the actual definition,''} and P7 (UXR) wondered how \textit{"how topic strength [..] is different from frequency.''.} Overall, participants wanted plain-language definitions, on-demand explanations, and evidence-backed justifications tied to each metric and assignment.

\subsubsection{Need to validate and Edit AI Outputs, Supporting Human Agency}

Participants wanted firm control over AI outputs—editing topic titles, reassigning or excluding quotes, and adjusting linked research objectives—to preserve interpretive flexibility, essential for their ability to review and approve the AI-generated analysis. As P12 (Lead UXR) said, they needed to \textit{``add and remove topics [..] edit [..] how we specifically categorize it [..] drag the topics around''} so related ideas can be merged.

They also asked for ways to challenge the AI. P18 (PM) \textit{``would love to [..] say you’re wrong [..] this quote [isn't] supporting this,''} and expected that feedback to shape subsequent suggestions. Several resisted premature commitment: for P16 (UX Designer), early-stage actions like “approve/hide” felt \textit{``very strong''} during an initial pass: \textit{``not sure if I will feel like to approve anything already.''}

However, control features must not erode trust. P16 (UX Designer) worried that a “Find similar quotes” button implied the system hadn’t actually surfaced everything, \textit{``“mak[ing] me think [..] you really didn’t go through all the quotes.''}


\subsubsection{Flexible Views and Visualizations to Support Individual Sense-making}
Participants wanted multiple, configurable ways to see the data. Many liked the bubble/Mural-style overview for quick spatial grouping because \textit{``how it is visually represented [..] would save me a lot of time''} (P15, UXR) and afforded the ability to \textit{``give it colors or labels''} to stay on top of what was said (P9, Innovation Designer). Some also wanted to \textit{``have everything displayed at all times [] and move them around [..] just like all of the murals of this world''} P9 (Innovation Designer), with basic zoom and \textit{``rearrange''} controls to fit their analytic style (P4, UXR).

Outliers were seen as trust signals, but views should offer choices. P12 (Lead UXR) said \textit{``I just want like a list [of outliers topics]''} and questioned the meaning for a spatial layout for outliers, while others argued \textit{``1000\% this is where all the signals are [..] the edge cases.''}

Beyond bubbles, participants asked to toggle between tabular, hierarchical, research-objective, participant/persona, and focused views: \textit{``a spreadsheet or like grid format is pretty easy for me to to digest''} (P12, Lead UXR); \textit{``expand out [to] full page or [..] because otherwise it's just going to be super annoying''} (P18, PM).
This reflects a desire to align views with individual roles and analysis practices — from broad overviews to deep dives




Filtering and slicing were essential for sense-making---by participant, research question, sentiment, frequency, custom tags, custom colors, or role-relevant facets. P9 (Innovation Designer) desired to \textit{``categorize everything in whatever way''} with a \textit{``customizable representation of the data.''} and P8 (Lead UXR) noted, that people need \textit{``something that's top down and then also something that's bottom up''}. Other emphasized tailoring to evolving interpretations because \textit{``sometimes is a very unlogical process''} and insights can \textit{``suddenly [..] come to the surface again''} P9 (Innovation Designer). A synchronized multi-view approach (spatial + list + hierarchy), with shared selections, fast zoom/pan, and one-click lenses, may best satisfy these varied analytical needs.

\subsubsection{Chat as Inquiry and Action Interface}
Participant found the chat interface mostly useful to ask questions about the data (like sophisticated search) - \textit{"It helps like you can ask a direct question about why something is the case that it is and stuff."} (P15, UXR), generate insights - \textit{" [...] we can ask the AI for suggestions to address the concerns raised".} (P5, PM), validate AI findings - \textit{"I can ask probing questions where I say maybe topic one and three is same or what do you think of this [...]".} (P6, UXR), or even take action -\textit{if I ask a question here right in the chat, maybe that should in itself like create a new topic or category or whatever and then fine tune or adjust this} (P5, PM). However, not all were comfortable with conversational interaction for analytical tasks: \textit{"I did also see the chat at first, but I've tried to ignore that. I think I just have bad feelings about chat bots."} (P21, UXR)
These findings suggest that chat works best as a context-aware tool for retrieval, explanation, and lightweight actions—valuable alongside, but not a replacement for, the primary analytic workspace.








\subsubsection{Collaboration and Sharing for Flexible Workflows}

Participants saw value in supporting not only solo analysis but also coordinated, multi-person workflows. They emphasized role-based access, clear ownership, moderation in group setting, and support for cross-team input when responsibilities are shared. As one lead UXR put it, \textit{``there should probably be [..] a captain or a lead [..]  otherwise things could get a little crazy''} (P12, Lead UXR). Collaboration needs also varied by team structure: \textit{``Co-analyze in a collaborative fashion? I think it depends [..] who's responsible [..] sometimes you don’t have a researcher''} (P4, UXR).

Participants also expressed openness to stakeholder input. PMs, Designers, Developers could validate emerging insights, \textit{``give a different perspective[s]''} or \textit{``flag things that seem out of place''} (P21, UXR). While an encouraged practice in our company, immediate debriefs after an interview are not common due to participation challenges (outlined in Q1). Using our tool for session debrief purposes, P1 (Senior UX Researcher) noted that \textit{``this would speed up the process and will be really valuable, especially if everyone's able to to have access''} and makes stakeholders \textit{``aware of the feedback.''}

These observations suggest that while collaboration is valued, effective AI-assisted group analysis requires structured roles, access controls, and support for layered perspectives, enabling researchers to guide the analysis while selectively involving others in meaningful ways.

\subsubsection{Summary RQ3}
Overall, most participants were excited about the tool we presented. We received more than 160 mentions of concrete feature ideas and design improvements, providing valuable input for future designs. Overall, the design probe revealed that UX researchers and PMs envision AI not as a replacement but rather as an assistant that can help with qualitative workflows by providing transparent analysis, fluid interaction styles, and human-in-the-loop control. They emphasized the importance of traceability, explainability, and evidence-based results, with UXRs prioritizing analytical depth, transparency, and accountability, and PMs focusing on speed, higher-level results, and reporting. Feedback also revealed the need for flexible interfaces that accommodate individual sense-making styles, role-based access, and support for collaborative workflows.









\section{Study 2: Methodology} \label{sec:study2}
Study 1 revealed that most user researchers experimented with GenAI using top-down approaches to thematic analysis of interview data, but they frequently encountered major drawbacks—missing themes and observations, inaccurate or irrelevant quotes, limited control over the process, and opaque reasoning behind the results.

In study 2 we sought to validate a bottom-up approach that can support human validation at every step. In the following experiment we focused only on the part related to extracting topics from interview statements and generating a coherent topic list for a single interview, to get an initial sense for the feasibility of this approach. Further steps are left to future research. Study 2 answers RQ4 of our investigation.

\subsection{System Design}\label{sec:system}
We present an approach for producing qualitative analysis of an interview transcript, with a LLM combined with a more traditional ML approach, inspired by \cite{Lam2024, CollabCoder2024}. 
Our system (see pseudo-code in Appendix~\ref{appendix:study2details}) accepts the following inputs: (1) an interview transcript, (2) a list of Research Objectives (ROs) that is used to prime the LLM  for the data extraction, (3) the maximum number of topics to be generated, and (4) a context size that determines how many conversational turns preceding the current statement should be included in the analysis. This approach can be divided into three stages: (1) \emph{Data Extraction}, (2) \emph{Clustering}, (3) \emph{Name Generation}.

In the \emph{Data Extraction} stage, we systematically parse the statements made by the interviewee, and prompt (see Appendix~\ref{appendix:prompt_topic_extraction}) an LLM to extract topics related to the ROs. Here (to the LLM and later to the SMEs) the topics were described as ``a higher-level theme, concept, or observation, and not just a keyword''. We did not opt for open ended extraction (as in \cite{Lam2024, CollabCoder2024}), since a guided extraction process produced more relevant Topics. After manual (experimenter) inspection of the topics generated by Llama-3-3-70b-instruct and GPT-OSS-120b-chat, we decided to use Llama-3-3-70b-instruct because it generated concise, code-like topics compared to GPT-OSS-120b-chat.
During the \emph{Clustering}, we generate embeddings of the topics and cluster them using  HDBSCAN \cite{campello2013density}. This approach was adopted from \cite{Lam2024} and is different from the LLM-based clustering available in \cite{CollabCoder2024}. Lastly, during the \emph{Name Generation} stage, we prompt an LLM to generate a name and summary (see Appendix~\ref{appendix:prompt_topic_cluster_name} and \ref{appendix:prompt_topic_cluster_summary}) that represent the topics in a cluster.

\subsection{Participants} 
To validate the output of our system, we 
recruited four subject matter experts (SMEs) from our research team. The sample consisted of 4 user experience researchers. 

\subsection{Data} 
We selected two transcripts (P4 and P9)  from Study 1, excluding design probe data due to poor LLM performance given the lack of context of the visuals.
P4 included 92 statements (37:44) and P9 had 129 statements (30:57). Both were chosen based on team feedback for clarity of articulation, to minimize issues with transcription quality. 
The anonymized data was converted to Excel  before processing.

\subsection{Procedure}\label{sec:study2_procedure}
We ran Study 2 with two conditions: Human-AI-Coding (HAIC) and Human-only-Coding (HoC). 
In the HoC condition, participants generated 0-5 topics for each statement, assigned each one a matching Research Objective (RO), and then clustered topics and named each cluster.

In the HAIC condition, participants evaluated the AI-generated analysis, focusing on the Topic, RO, and Topic Cluster Name (TCN). 
Participants were asked to answer the following questions on a scale of 1 (low score, bad match) to 5 (high score, great match):

\begin{itemize}
    \item Q1: The ``Topic'' is a good match for the ``Statement''. 
    \item Q2: The ``Research Objective'' is a good match for the statement. 
    \item Q3: The ``Topic'' is a good match for the cluster ``Topic Cluster Name''. 
    \item Q4: The ``Topic Cluster Name'' is a good representative title for the topics in the cluster. This is assigned at cluster level. 
    \item Q5: The ``Topic Cluster Summary'' is a good representative description for the topics in the cluster. This is assigned at cluster level. 
\end{itemize}

We also asked to rate each AI outcome with ``Accept AI Analysis? (Yes/No)''. If \textit{No}, they had to provide revised values for the Topic, RO, or TCN in dedicated columns, or enter new values if needed. SMEs also provided written feedback after completing their task.

We asked each SMEs to evaluate an AI-generated analysis in the HAIC condition and to complete a manual analysis in the HoC condition. The study was administered in two rounds. In each round, groups of two participants were independently assigned to one condition, same data (either P4 or P9). In the second round, the group of participants switched condition using the other data (either P4 or P9).
We ensured that the Study 2 participants  were not the original analysts from Study 1 for those transcript and that each participant had a unique condition-transcript pairing across rounds.

\section{Study 2: Results}
\label{sec:study2results}

The following results address \textbf{RQ4 - How feasible is a bottom-up LLM-based approach in performing thematic analysis similar to traditional human coding methods??}

\subsection{How well did the AI perform in analyzing qualitative data?}

\begin{figure}[htbp]
  \centering
  \begin{subfigure}[t]{0.48\textwidth}
    \centering
    \includegraphics[width=\linewidth]{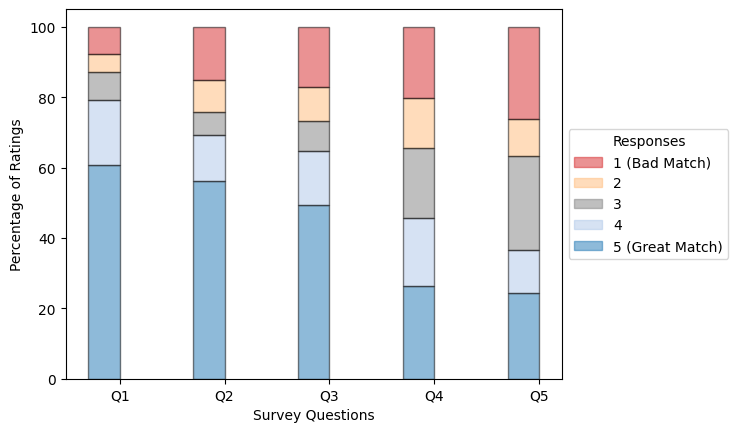}
    \caption{Survey results for AI analysis, user ratings for Q1 (Topic-Statement match), Q2 (RO-Statement match), Q3 (Topic-TCN match), Q4 (TCN-all Topics match), Q5 (TCS-all Topics match), rated on a scale of 1 (Bad match) to 5 (Great match).}
    \Description{Survey results for AI analysis, user ratings (from left to right) for Q1 (Topic-Statement match), Q2 (RO-Statement match), Q3 (Topic-TCN match), Q4 (TCN-all Topics match), Q5 (TCS-all Topics match), rated on a scale of 1 (Bad match) to 5 (Great match). The results show the number of high ratings decreasing and low ratings increasing from left to right.}
    \label{fig:survey_results_all}
  \end{subfigure}
  \hfill
  \begin{subfigure}[t]{0.48\textwidth}
    \centering
    \includegraphics[width=\linewidth]{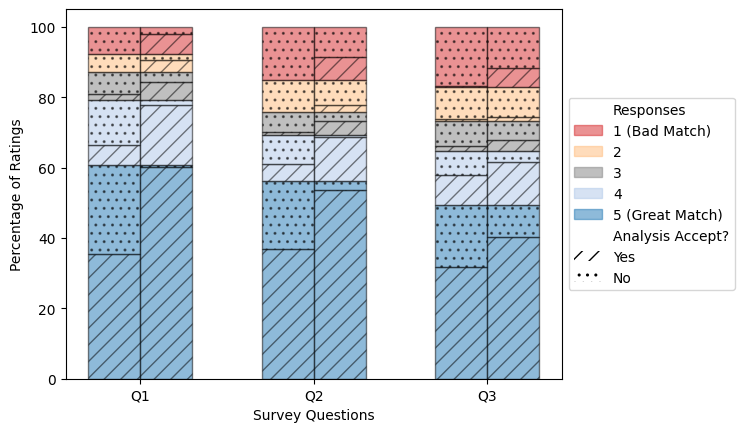}
    \caption{Q1–Q3 paired bar-chart of ratings split by percentage of samples with AI analysis accepted with yes/no. Left bar shows users' direct ratings. Right bar shows users' adjusted rating, accounting for value changes to respective values addressed by questions.}
    \Description{Q1–Q3 paired bar-chart of ratings split by percentage of samples with AI analysis accepted with yes/no. Left bar shows users' direct ratings. Right bar shows users' adjusted rating, accounting for value changes to respective values addressed by questions. The trend shows the number of acceptances of AI analysis increasing when we take into account if the user changed the value or not.}
    \label{fig:survey_results_q1q2q3_split}
  \end{subfigure}
  \caption{Combined survey results: (a) overall, (b) Q1–Q3 detail.}
  \Description{Two subfigures side by side showing overall survey results and a split view for Q1–Q3.}
\end{figure}

The number of statements analyzed by AI across the two transcripts was 248. Since each transcript was rated by two users, we obtained 494 ratings. 
Figure~\ref{fig:survey_results_all} presents the SME ratings of the AI-generated analysis. Results show that almost 80\% of the AI-generated topics were rated a good or great match (Q1) and almost 75\% of the assigned research objectives were rated a good or great match (Q2).
However, once algorithmic clustering was applied, ratings dropped slightly for topic-cluster alignment (Q3)
with further decline in ratings for the 
the LLM (Llama-3-3-70b-instruct) generated cluster name and summary (Q4 and Q5). 

Figure~\ref{fig:survey_results_q1q2q3_split} provides a more detailed view, in a paired bar chart, where in each pair, the bar on the left represents users' direct rating for the questions Q1, Q2 and Q3, split by the yes/no value of accepting the AI analysis. Rejection was defined as
at least one of the values in the triplet (Topic, RO, TCN) was not acceptable to the participant ('No'). 
we adjusted the acceptance value based on whether or not  the user had provided a new value for Topic (Q1), RO (Q2), and TCN (Q3). 
These adjustments are reflected in the plot in each bar on the right side of each pair. We can observe the number of responses increase for accepting the analysis for the respective items.

\subsection{SME feedback on AI analysis}
SME feedback indicated that the generated topics and assigned ROs were good, although on some occasions the topics focused on non-important parts of the statement, and the topics were often higher level than desired. Regarding clustering SME2 said \textit{``I had really good clusters but some clusters would need to be divided into two [..] it will require more information when clustering.''} which resonated with other SMEs as well. A couple of SMEs mentioned how some topics were often assigned to incorrect clusters.

\subsection{What is the difference between partnering with an AI analyst vs human analyst}

To explore how AI compares to a human analyst in collaborative coding,  
we compared topics generated by two humans (80 comparisons) and by a human-AI (172 comparisons) pair. 
Since topic generation was free-form and not based on a shared codebook, we prompted (see Appendix~\ref{appendix:prompt_topic_list_compare}) an LLM (Llama-3-3-70b-instruct) to asses semantic similarity.
The LLM identified overlapping (similar) and unique topics across each pair, allowing us to 
compute the set intersection value, while that along with the unique values can be used to understand the set union value. Using these, we compute the Jaccard Index (see Equation~\ref{eq:jacc_idx_eq}) as a similarity metric to compare the human-human scores to human-AI scores.

\begin{equation}\label{eq:jacc_idx_eq}
    J(A, B) = \frac{A \cap B}{A \cup B} = \frac{A \cap B}{(A \cap B) + (A - B) + (B - A)}
\end{equation}

A Welch's \textit{t}-test (due to unequal sample sizes) comparing Human-Human (mean: 0.51, std. dev.: 0.36) and Human-AI (mean: 0.41, std. dev.: 0.31) Jaccard scores showed significant statistical difference (t-statistic: 2.23, p-value: 0.027). Across the 4 participants we found 12 unique statements for which the human had not assigned any topics, but the AI had assigned additional 10 topics in P9 (2 accepted by SME3 and 6 by SME4) and 21 topics in P4 (8 accepted by SME1 and 16 by SME2). 

\subsection{Summary RQ4}
Overall, the LLM was able to generate Topics and assign ROs that participants validated as good or great matches in most cases. 
Algorithmic clustering also received favorable ratings. Although
some items were rated lower, the majority of the AI-generated analysis was accepted by the users. Cluster name and summary generation performed less well, often due to topics being misassigned to clusters or cluster abstractions being too broad. Some of these issues can attributed to the HDBSCAN algorithm's management of noisy samples\footnote{https://scikit-learn.org/stable/modules/generated/sklearn.cluster.HDBSCAN.html}. 
When comparing inter-annotator agreement, significant differences were found between human-human and human-AI analysis, suggesting that partnering with an AI may be different compared to a second analyst in the setting, with AI contributing ideas not entirely similar to your own. However, annotation styles and preferences between humans are widely different. Our sample size is too small to draw final conclusions. Topics were generally rated well by our SMEs and based their feedback, they can be further improved by being more detailed. Overall, the above results are a promising direction for implementing a bottom-up approach while some of the clustering challenges will still need to be addressed in future work.




\section{Discussion and Design Implications}

Our findings paint a complex picture of how GenAI is beginning to reshape qualitative research practices in software product development. Guided by our four research questions, we uncovered differences in workflows, expectations, and experiences with AI across user experience researchers, product managers, and designers.

RQ1 showed that people approach user research in different ways: UXRs often follow more structured thematic analysis methods, while PMs prioritize synthesizing information for stakeholder communication. These distinct roles shape different expectations for AI, depth and interpretability for UXRs versus speed and actionable summaries for PMs.

RQ2 revealed differing attitudes toward GenAI between PMs and UXRs. While both saw potential to streamline research, PMs were more optimistic, focusing on speed and efficiency. UXRs were more cautious, stressing the need for human validation, interpretive nuance, and transparency. Their concerns stemmed from a deeper engagement with user data and fears of missing critical insights. Both groups expected AI to play a larger role in research workflows in the future, but with different needs, levels of trust, and enthusiasm. 

RQ3 explored participants’ expectations for AI-powered tools through a design probe. Most were excited about the tool, with more positive than negative reactions and over 160 suggestions for improvements. Participants saw AI as a helpful assistant—not a replacement—for qualitative workflows, emphasizing the need for transparency, traceability, and human-in-the-loop control. The results underscored the importance of flexible, role-aware interfaces that support individual sense-making and collaboration.

RQ4 assessed the feasibility of a bottom-up AI analysis pipeline. In a validation study, we found the AI-assigned topics and research objectives to be mostly agreed-upon and useful, though some shortcomings appeared in how clusters were named or summarized. Importantly, participants accepted the majority of AI-generated analyses when allowed to review and refine them.

Our results point to key design opportunities, 
which we articulate below through a set of implications for building trustworthy, human-centered AI tools for qualitative analysis.

\subsection{Make provenance and explainability of the AI generated analysis the backbone of qualitative analysis}
In keeping with \citet{Amershi2014}, we found that trust in AI-assisted qualitative analysis depends on the ability to trace every suggestion—whether a theme, topic, or insight—back to its supporting evidence and the original transcript context. To support this, tools should default to bidirectional links across all analytical artifacts (quotes, codes, clusters, summaries) and display rationales for inclusion and transformation. These rationales should be concrete and inspectable, for example, by highlighting salient terms, showing semantic neighbors, or visualizing dispersion across participants. Presenting counter-evidence, including quotes that challenge the current grouping, can further strengthen analytical rigor. Finally, approval processes could be gated: brief reviews or spot checks tied to raw evidence can reinforce researcher control and accountability.

\subsection{Design for human agency: validate, edit, and challenge}
\citet{goyanes2025thematic} and \citet{naeem2025thematic} explored multi-step analytic protocols in which a human could intervene in the LLM's series of steps. Our participants confirmed their human-in-control configurations. We found strong evidence that participants desire to maintain control over AI-generated analysis. Analytics tools should make AI-generated artifacts editable in place, such as renaming or merging themes or topics, reassigning or excluding quotes, and annotating decisions. The downstream implications of such edits on the overall analysis should be made visible to foster informed decision-making. Challenge mechanisms should be lightweight but meaningful (e.g., “reject with reason”) and could feed into a correction memory that can inform future suggestions. To support human judgment, systems should surface relevant contextual metadata, for example, participant type, product area, or scenario, alongside themes and quotes. To lower the cognitive overhead of maintaining human oversight, tools could include low-effort scaffolds, such as, for example, inline edit controls, quick-reject tags, or alternative suggestions, that make it feasible to challenge AI outputs, even under time pressure.

\subsection{Support hybrid top-down and bottom-up analysis workflows}

Our results show that some participants employed a top-down approach while others used AI bottom-up, generating preliminary codes or quote-level labels which were then grouped into broader themes. As thematic analysis has evolved \cite{braun2023toward}, both approaches are used, and our findings reflect the diversity among UXRs as well as the diversity among accepted practices. Top-down approaches, such as prompting AI to suggest preliminary themes, offer a directional overview, and may serve as a starting point rather than a final output. Participants acknowledged that such summaries may lack accuracy but still function as useful “headlights” for early sense-making. Bottom-up approaches allow humans to stay closely engaged with the data by validating quote-level codes and progressively grouping them into broader themes. Our experiments in Study 2 demonstrated the viability of this human-in-the-loop process for generating trustworthy outcomes. We envision tools  that are flexible enough to support both ways of approaching the data, enabling researchers to generate theme drafts with supporting evidence (top-down), while also surfacing quote-level code suggestions (bottom-up). Since qualitative analysis is an iterative process, both approaches might be helpful in different stages of the analysis process. The success of such hybrid workflows depends on minimizing the cognitive and interactional burden of editing.

\subsection{Integrate chat as a bridge, not a destination}
Most participants had previously used chat as a standalone feature to experiment with GenAI with mixed results, mostly related to issues with context-size and manual effort of organizing prompts, input data and results. In fact, in our design probe participants did not universally endorse chat as a central interface for analysis. However, many acknowledged its value for data retrieval from transcripts and other artifacts, explanations, probing questions, insight generation, and quick actions. A tight integration of chat in the analytics environment would release the burden of prompt, data, and results management by situating interactions within the context of the current analysis state. Inspired by popular vibe-coding environments (e.g., Windsurf\footnote{https://windsurf.com/} or Cursor\footnote{https://cursor.com/en}), this vision positions chat as a fluid bridge between human intent and AI-powered tooling, i.e. it would preserve more expressive or visual modes of the analysis.


\subsection{Replace opaque metrics with meaningful signals that establish trust}

According to our findings, PMs were generally satisfied with high-level summaries and focus on key topics, while UXRs emphasized the need for more nuanced and interpretable metrics. For them metrics must go beyond surface-level insights to support  deeper understanding, contextual grounding and defensible insights.Thereby, we offer important updates to \citet{Sankaranarayanan2025}'s proposed criteria of robustness, transparency, and validity. Those concepts will be useful to industry workers only to the extent that they are translated into work-oriented metrics that are already in practical use. 

Both PMs and UXRs 
expressed concerns about AI trustworthiness and stressed the importance of evidence to justify results. Participants widely understood simple metrics like frequency, but reported confusion around less transparent indicators such as “coherence,” “strength,” or color-based encodings. To build trust, we suggest replacing opaque or model-internal metrics with signals that are meaningful within qualitative analysis workflows, such as, for example, dispersion across participants, rarity (outliers), sentiment, novelty, alignment with research questions, and stability across multiple iterations. Additionally, we believe that representative evidence and counter-examples should help support interpretation and judgment of metrics.


\subsection{Role-adaptive interaction and access}
\label{sec:role-adaptive}
Our findings highlight the importance of designing AI-assisted qualitative analysis tools that adapt to the diverse needs of different roles \cite{jameson2007adaptive}. PMs typically seek concise, shareable outcomes to inform decision-making, while UXRs require in-depth access to data, editability, and auditability to ensure analytic rigor. Role-adaptive interaction could take the form of different interface modes, for example, a research mode with raw data views, coding controls, and change history, and a stakeholder mode that surfaces top findings with embedded citations and read-only summaries. Customizable dashboards and access controls would ensure that users see content appropriate to their role, task, and responsibilities. Permissions (e.g., lead, editor, commenter, read-only) and approval gates for risky edits can further support collaboration and data integrity. Additionally, features like “render to deck” and shareable snapshots could translate the current analytic state into presentation-ready content without cutting the links back to the source data, which maintains traceability across roles.


\subsection{Support individual sense-making styles with multiple, synchronized views}
Many participants indicated that no single view supports all analytics strategies or users' preferences with some having strong preferences for either visual or tabular representations. Hence, an analytics tool should be a multi-view environment that synchronizes between spatial, list-based, and hierarchical representations of the analytics state and data to accommodate diverse sense-making styles (e.g., \cite{li2018design}). Synchronizing selections, highlights, and filters across these views can preserve analytic continuity. Features such as fast zoom, a mini-map for context, and one-click lenses to reconfigure the corpus by research question, participant/persona, sentiment, or frequency can support flexible exploration. 

We heard that outliers are really important. They should be treated as first-class entities by displaying them in a dedicated outlier space with contextual explanations for their status (which may change over the course of the analysis), while also appearing in relevant theme or topic groupings.

\subsection{AI-Augmented Debriefs}
Our findings suggest that post-interview debriefs can accelerate insight generation by helping teams align on what they heard and surface key evidence while the content of the interview is still fresh in their minds. Despite being encouraged, participants reported that they are not done consistently. We see an opportunity for AI to enhance this practice by supporting real-time capture, coordination, and synthesis. A live or debrief mode in the tool could capture transcripts in real-time with speaker attribution, surface tentative quotes and topics. A debrief assistant could support deliberation of the tentative results and surface AI-role-tagged notes (reflecting a PM or Design perspective) anchored to evidence to be reviewed and merged later. While we anticipate that a live approach may speed up results, one would have to overcome the hesitation we observed in introducing AI results live without prior review.


\subsection{Collaboration Within and Between Roles}

Earlier, we noted the collaborative nature of UX research \cite{braun2022conceptual, braun2023toward, bowman2023using}. Participants told us repeatedly of their need to combine perspectives and ideas during debriefs and in more formal meetings. UXRs often needed to discuss topics in detail within a Team setting, but then provide much more summarized and high-level impactful materials to PMs for presentation in their Circles with other departments. We propose that UXRs may need features to curate particular thematic analysis findings for use by a PM, choose the adaptive re-framing of those findings into a summarization for a specified audience (see Section \ref{sec:role-adaptive}), and then notify the PM of the availability of these client-customized materials.

\section{Limitations}
We acknowledge several imitations in our research. All of our participants worked in the same global technology company, across six countries, which may not reflect the reality for other organizations, specifically practices may differ across companies and different domains and products, although we believe that the basic principles of qualitative analysis apply more broadly in product development. Additionally our participant sample was composed of UXRs, PMs, and designers, constrained to a specific subset of stakeholders in the software development life cycle, excluding perspectives of developers, clients, sales, marketing, and executives. Future work could expand the participant base to include a more diverse range of roles, organizations, and product categories. We also validated the bottom-up approach with a small number of internal experts and only two transcripts, which limits generalizability. More work is needed to cover additional aspects of bottom-up approaches (e.g. grouping, code book management) and will be covered in future work as the goal here was to just get an initial feasibility understanding. Study 1 relies on self-report and a design probe rather than in-the-wild tool use, so stated practices and preferences may not fully predict behavior under real constraints. Study 2’s pipeline evaluation is model- and parameter-dependent (prompting, embeddings, HDBSCAN), and cluster naming/summarization quality can be sensitive to those choices.

\section{Conclusion}
We conducted two exploratory studies to examined how two major stakeholder groups---UXRs and PMs---approach qualitative analysis (RQ1). We explored current uses, perceived challenges, and future opportunities for integrating GenAI into qualitative analysis workflows (RQ2), and investigated the envisioned user experiences and interaction patterns UX professionals anticipate for effective GenAI integration through a design probe (RQ3). Finally, we validated a bottom-up, LLM-based approach to thematic analysis (RQ4). Our research aims to illuminate both prevailing practices and speculative futures, offering grounded insights into how GenAI might reshape qualitative research in software development contexts. 

GenAI lowers the barriers to participation in user research---but in doing so, it amplifies the risk of misuse. Concerns are that non-experts may rely on model outputs without the methodological grounding to assess their validity, or selectively use findings to support biased narratives. The solution is not to restrict access, but to embed accountability within the tools themselves, such as evidence-linked outputs, checkpoint reviews before publishing results, role-sensitive interaction modes, or alternative interpretations. While models may generate suggestions, systems must provide justification, and humans must remain the interpretive authority. We hope the eight design opportunities outlined in this paper offer concrete 
guidance for building AI systems that not only accelerate qualitative analysis but also preserve the rigor, transparency, and critical judgments that UXRs value. 

\bibliographystyle{ACM-Reference-Format}
\bibliography{references}

\newpage



\appendix

\section{Interview Guide}
\label{appendix:guide}

\textbf{[1] Background}

\begin{enumerate}
    \item What types of user studies do you typically conduct?
    \item What is your experience with thematic analysis?
    \item IF PRODUCT MANAGER: Why do you think it’s important to run user studies?
\end{enumerate}

\noindent
\textbf{[02] As-Is Scenario}
\\If the interviewee was a product manager, the following questions were adjusted by the moderator to account for the the perspective of a product manager as appropriate.

\begin{enumerate}
    \item What types of user studies do you typically conduct (e.g., usability, generative, evaluative)?

    \item How do you get access to the information (sources) you need to plan the user studies? What background information of the domain of the study/product is important and where do you get it from?

    \item How long does it take to run a study and analyze the data collected?

    \item When running your analysis, what are the main criteria you use to identify relevant topics from the interview transcripts?

    \item How do you manage your analysis process (including themes, observations, findings, topics)?

    \item How do you arrive at the themes? Where do you capture themes? After each interview or after more interviews? How many usually?

    \item How do you revise the themes?

    \item What tools do you use for analyzing the data collected during user experiments, and debriefing sessions? How have you used these tools?

    \item What are major challenges / obstacles for you in conducting thematic analysis?

    \item How do you think thematic analysis could be accelerated without sacrificing the integrity and quality of the analysis?
\end{enumerate}

\noindent
\textbf{[03] - Collaboration and sharing the results}

\begin{enumerate}
    \item How do the stakeholders and team (PM, designers, developers) participate in user research? In which phases? Please could you detail the dynamics of those activities?
    
    \item Do you run debrief sessions after each interview? With whom? How does it work?\\
    If not, why not?

    \item What are the major benefits of debrief sessions? What are the biggest challenges?

   \item OPTIONAL: What’s the output and how do you use it?

    \item OPTIONAL: How do you communicate the findings of user studies to the rest of the team? Do you have to show preliminary results?
    
\end{enumerate}

\noindent
\textbf{[04] - Perceptions of tools and AI-assistance for thematic analysis}

\begin{enumerate}
    \item Have you used AI in your workflow? If yes, how? Was it effective?

    \item In your opinion, how should AI not be used in thematic analysis? Do you trust it? Why? Why not? (What is needed to have trust?)

    \item OPTIONAL: How has your tool usage changed from the early introduction of GenAI to now?

    \item OPTIONAL: How do you think AI will change the nature of user studies and qualitative data analysis in about 2 to 5 years?
    
\end{enumerate}

\noindent
\textbf{[05] - Design probe}
\begin{enumerate}
    \item Discuss Screenshot 1 (see Appendix Figure \ref{fig:screenshot1})
    \item Discuss Screenshot 2 (see Figure \ref{fig:teaser})
    \item Discuss Screenshot 3 (see Appendix Figure \ref{fig:screenshot3})
    \item Discuss Screenshot 4 (see Appendix Figure \ref{fig:screenshot4})
    \item Discuss Screenshot 5 (see Appendix Figure \ref{fig:screenshot5})
\end{enumerate}

\noindent
\textbf{[06] - Understanding Adoption}

\begin{enumerate}
    \item Is there any feature in this tool that reminds you about other tools you used before?
    \item Do you think you need to do any preparation before using this tool? Which one?
    \item Would this tool be useful to use during the debrief session after the interview? Explain!
    \item Would you use it together with your team or on your own? (collaboratively)
    \item What concerns do you have about this tool?
    \item OPTIONAL: What happens in your opinion, after you use this tool? What is the next step? (check if they would like to download the info to use in another place)
\end{enumerate}

\section{Design Probe}

\begin{figure}[htbp]
    \centering
    \includegraphics[width=0.9\linewidth]{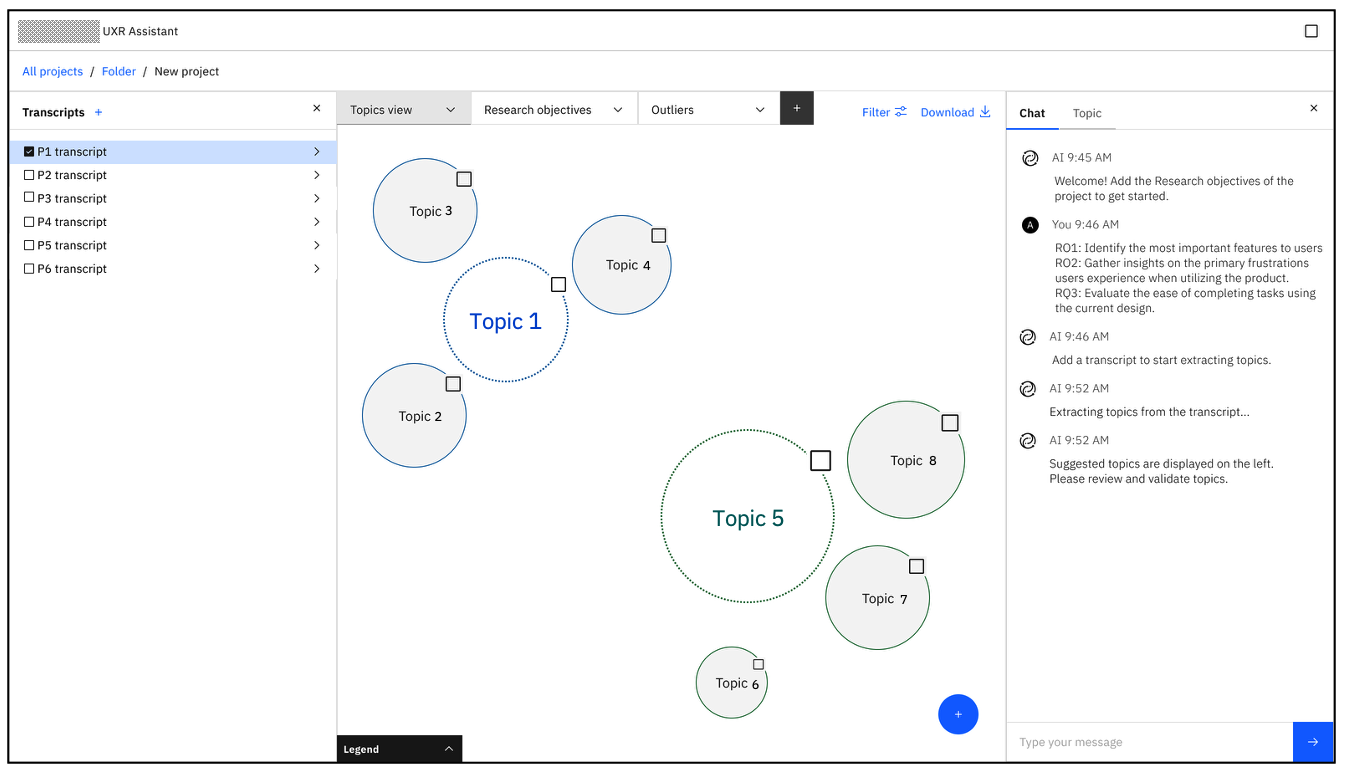}
    \caption{Mock-up of AI-extracted topics with the initiation flow shown in the chat. In the chat, the user provides research objectives of their study, and the AI requests transcripts to be uploaded. After the user adds the transcripts, the AI displays suggested topics and asks them to review and validate.}
    \Description{A mock-up showing a view of AI-extracted topics. The interface is split into three side-by-side panels. The left panel shows a list of study transcripts uploaded by the user -- there are six total, and ``P1 transcript'' is selected. The middle panel shows AI-extracted topics as circles with different sizes and outline colors. These circles are clustered into two groups with a central circle that represents the parent topic and smaller circles surrounding it that represent sub-topics. The right panel shows a chat interface with a conversation between the AI and user. The AI asks for research objectives, the user shares those objectives, the AI asks for transcripts, then informs the user that suggested topics have been extracted and displayed.}
    \label{fig:screenshot1}
\end{figure}

\begin{figure}[htbp]
    \centering
    \includegraphics[width=0.9\linewidth]{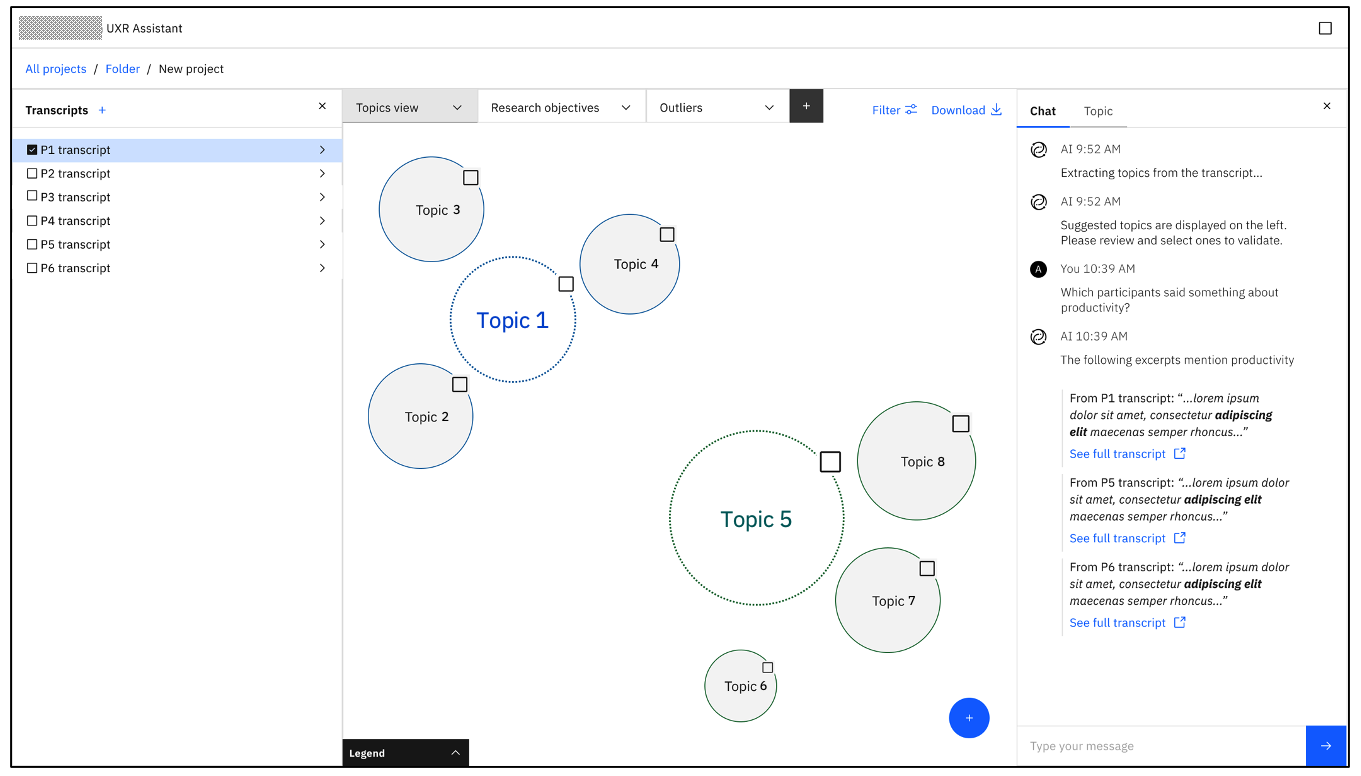}
    \caption{Mock-up of a chat flow in which a user requests the AI for insights. In the chat panel, the user asks, "Which participants said something about productivity?" The AI responds with three excerpts that mention productivity, along with the option to see the full transcript for each one.}
    \Description{A mock-up showing a view of AI-extracted topics. The interface is split into three side-by-side panels. The left panel shows a list of study transcripts uploaded by the user -- there are six total, and transcripts for P1, P3, P5, and P6 are selected. The middle panel shows AI-extracted topics as circles with different sizes and outline colors. These circles are clustered into two groups with a central circle that represents the parent topic and smaller circles surrounding it that represent sub-topics. The right panel shows a chat interface with a conversation between the AI and user. The user asks, "Which participants said something about productivity?" The AI responds with three excerpts that mention productivity, along links to see the full transcript for each one.}
    \label{fig:screenshot3}
\end{figure}

\begin{figure}[htbp]
    \centering
    \includegraphics[width=0.9\linewidth]{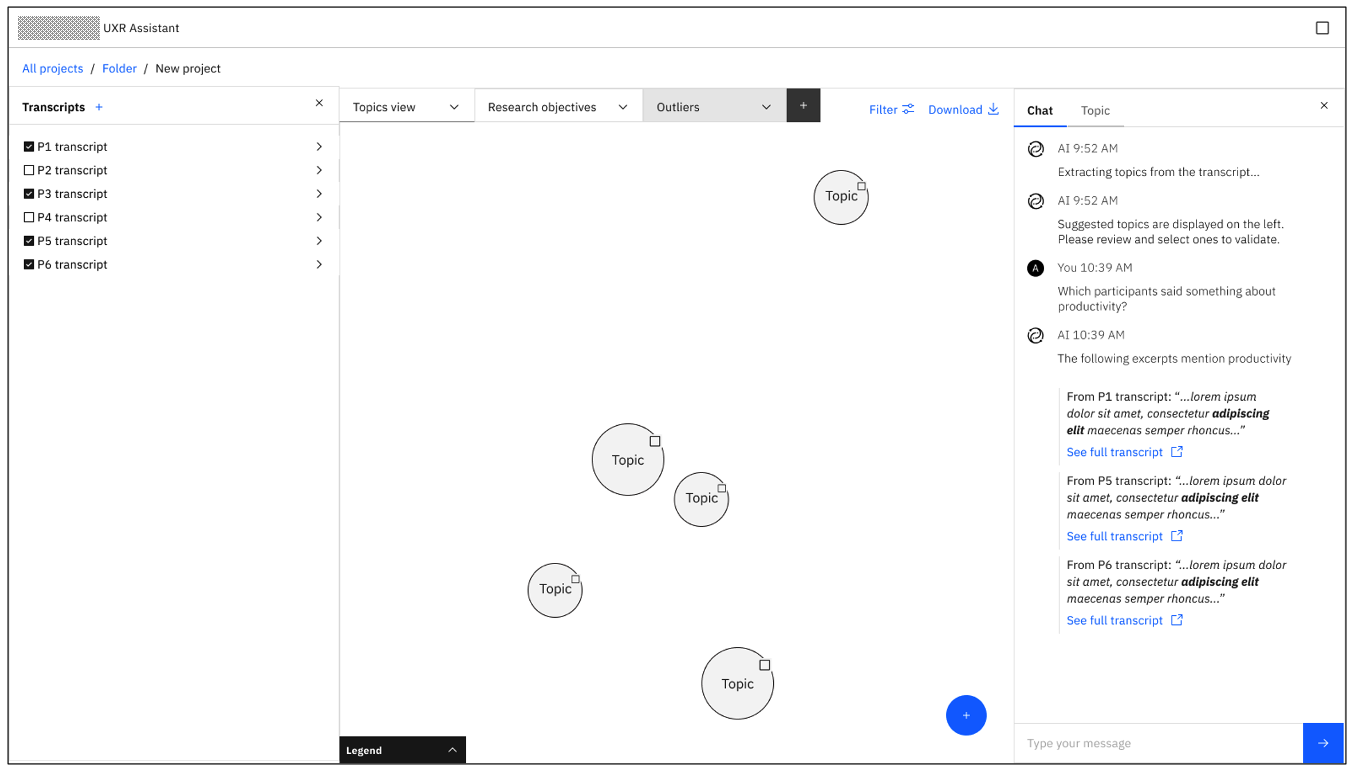}
    \caption{Mock-up showing outliers in AI-extracted topics. Outliers are displayed as small, disconnected circles, representing the smaller amount of data backing them up and their disconnect with topic clusters.}
    \Description{A mock-up showing a view of AI-extracted topics. The interface is split into three side-by-side panels. The left panel shows a list of study transcripts uploaded by the user -- there are six total, and transcripts for P1, P3, P5, and P6 are selected. The middle panel shows AI-extracted outliers as small circles with different sizes. These circles are scattered around the view. The right panel shows a chat interface with a conversation between the AI and user.}
    \label{fig:screenshot4}
\end{figure}

\begin{figure}[htbp]
    \centering
    \includegraphics[width=0.9\linewidth]{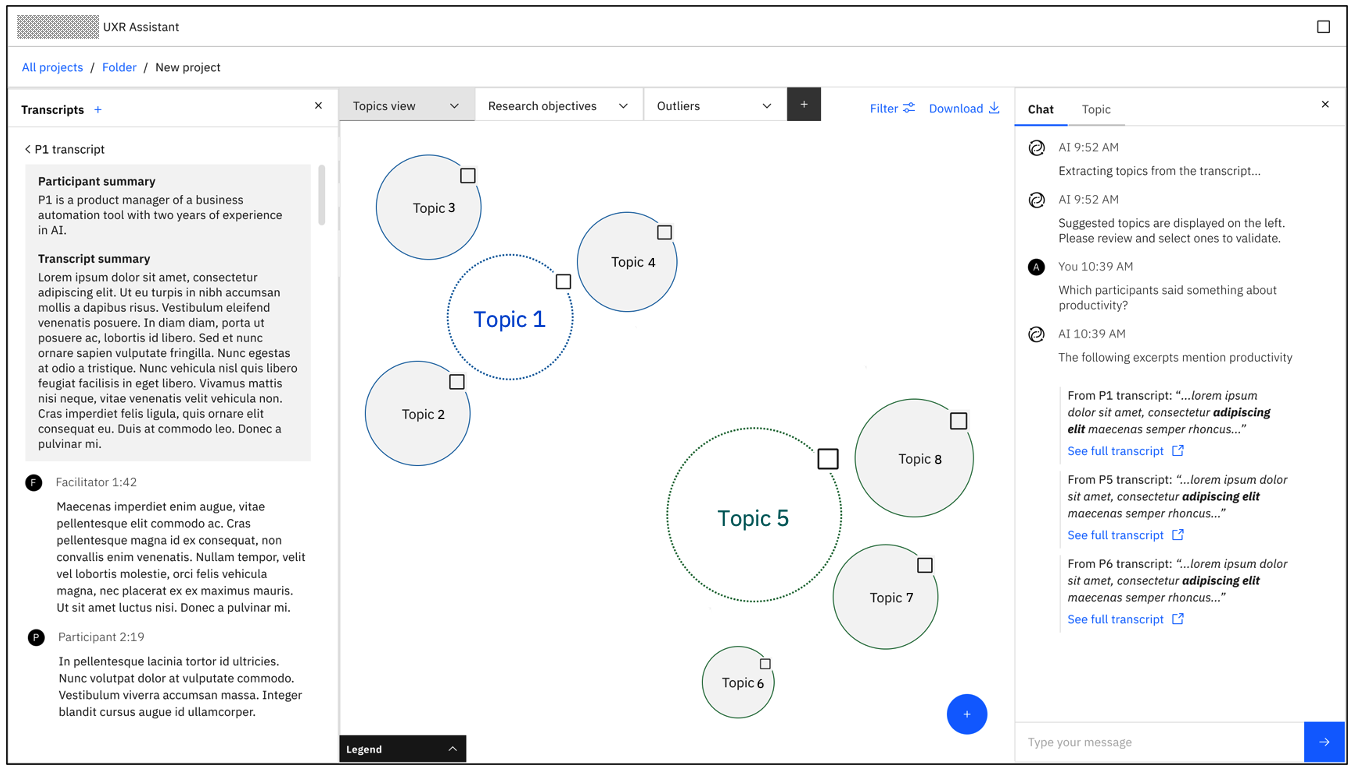}
    \caption{Mock-up showing the transcript view for one participant. In the transcripts panel on the left, P1 transcript has been selected. The panel displays a participant summary, transcript summary, and the full transcript below.}
    \Description{A mock-up showing a view of AI-extracted topics. The interface is split into three side-by-side panels. The left panel shows a the detailed view for one transcript, with a participant summary and transcript summary with placeholder text in a gray box at the top. Below the summary box is the full transcript between a facilitator and participant with placeholder text. The middle panel shows AI-extracted topics and clusters as groups of circles with different sizes and outline colors. The right panel shows a chat interface with a conversation between the AI and user.}
    \label{fig:screenshot5}
\end{figure}

\clearpage
\section{Codebook}
\begin{table*}[htbp!]
  \caption{Codebook}
  \label{apendix:codebook_1}
  \resizebox{\textwidth}{!}{%
  \begin{tabular}{p{2cm} p{2cm} p{3cm} p{5cm}}
    \toprule
    \textbf{Theme} & \textbf{Sub-Theme} & \textbf{Code(s)} & \textbf{Operational Definition} \\
    \midrule
    \multirow{2}{1pt}{Roles and Responsibilities} 
        & Role & Role - [UX Researcher, Product Manager, Designer, Senior UX Researcher, …] & The official job title or position held by the participant (e.g., UX Researcher, Product Manager). \\
        \cmidrule{2-4}
        & Tasks in Role & Task - [Speculative Design, market research, strategy, teaching new methods, informal interviews] & Specific activities or duties the participant performs as part of their role (e.g., speculative design, market research, etc.). \\
        \cmidrule{2-4}
        & Multi Role & Task - Multi-Role - [list] & When a participant has transitioned across multiple roles (e.g., design, PM, UX research), or holds overlapping responsibilities. \\
        \cmidrule{2-4}
        & \multirow{2}{1pt}{Experience TA} 
            & Level - [High/Medium/Low] & Participant’s self-described expertise in thematic analysis/qualitative analysis, categorized as high, medium, or low. \\
            \cmidrule{3-4}
            & & Years - [Years] & Number of years of experience in thematic/qualitative analysis explicitly stated by the participant. \\
            \cmidrule{3-4}
            & & Methods - [e.g. card-sorting etc. enter here] & Techniques or methodological approaches used by the participant in thematic/qualitative analysis. \\
            \cmidrule{2-4}
        & Team & Team - [name] & The named group or organizational unit the participant is affiliated with. \\
        \cmidrule{2-4}
        & Product & Product - [name] & The specific product(s) the participant is working on or referencing. \\
    \midrule
    \multirow{2}{1pt}{Research Process and Workflow}    
        & \multirow{2}{1pt}{Phases} 
            & Phases - Discovery & Activities during the discovery phase of a product. User studies in this phase tend to be more generative in nature. \\    
            \cmidrule{3-4}
            & & Phases - Planning & Activities involved in preparing for a study, including defining the scope, research questions, study design, logistics, and coordination of the research process. \\
            \cmidrule{3-4}
    \end{tabular}
    }
 \end{table*}
\clearpage
\begin{table*}[htbp!]
\centering
  \caption{Codebook (continued)}
  \label{apendix:codebook_2}
  \resizebox{\textwidth}{!}{%
  \begin{tabular}{p{2cm} p{2cm} p{3cm} p{5cm}}
    \toprule
    \textbf{Theme} & \textbf{Sub-Theme} & \textbf{Code(s)} & \textbf{Operational Definition} \\
    \midrule
    \multirow{2}{1pt}{Research Process and Workflow}    
        & \multirow{2}{1pt}{Phases} 
            & Phases - Recruiting & Actions and strategies used to identify, screen, and secure participants for the study, ensuring alignment with the research goals and target user profiles. \\
            \cmidrule{3-4}
            & & Phases - Data Collection & The processes and methods used to gather data from participants (e.g., interviews, surveys, usability testing, observations) during the execution of the study. \\
            \cmidrule{3-4}
            & & Phases - Synthesis & Analytical activities focused on reviewing, organizing, and interpreting collected data to generate patterns, insights, and preliminary findings. \\
            \cmidrule{3-4}
            & & Phases - Results Presentation & The preparation and communication of research findings to stakeholders through presentations, reports, or other deliverables to inform decision-making. \\
            \cmidrule{2-4}
        & Method & Method – [method name] & Specific research method(s) used (e.g., usability test, interviews, informal interviews, surveys, customer research). \\
        \cmidrule{2-4}
        & \multirow{2}{1pt}{Research Timeframes} 
            & Short-Term Product Needs & Research conducted to address immediate product decisions. \\    
            \cmidrule{3-4}
            & & Long-term Product Needs & Research conducted addressing future or strategic product directions. \\
            \cmidrule{2-4}
        & \multirow{2}{1pt}{Research Goals} 
            & Defining \& Aligning & Setting up the goals and making sure they are aligned including research plans, etc. \\    
            \cmidrule{3-4}
            & & Designing the Future & Research goals aimed at envisioning or shaping long-term product or user experiences. \\
            \cmidrule{3-4}
            & & General & Examples or descriptions of studies and goals, i.e. what the team is trying to accomplish by doing them. \\
            \cmidrule{2-4}
    \end{tabular}
    }
 \end{table*}
\clearpage
\begin{table*}[htbp!]
\centering
  \caption{Codebook (continued)}
  \label{apendix:codebook_3}
  \resizebox{\textwidth}{!}{%
  \begin{tabular}{p{2cm} p{2cm} p{3cm} p{5cm}}
    \toprule
    \textbf{Theme} & \textbf{Sub-Theme} & \textbf{Code(s)} & \textbf{Operational Definition} \\
    \midrule
    \multirow{2}{1pt}{Research Process and Workflow}    
        & Process Challenges & Challenge - [e.g. Lack of long-term research, Limited time for knowledge transfer, Duplication of work, Burn-out, Timeframes, Insufficient Time, Information Scatter, etc.] & Challenges, difficulties, painpoints experienced during the process of user research. \\
        \cmidrule{2-4}
        & Study Type & Type - [e,g, Evaluative, Generative, …] & \textit{Evaluative:} Research focused on assessing existing designs, products, or concepts.
\textit{Generative:} Research aimed at discovering new ideas, needs, or opportunities. \\
        \cmidrule{2-4}
        & Study Duration &Duration - [enter numbers and specify] & Captures mentions of study duration. \\
        \cmidrule{2-4}
        & Study Size &Size - [number of participants] & Captures mentions of study size. \\
        \cmidrule{2-4}
        & Product Stage &Stage Name - [early concept, prototype launch, mature product, or anything that fits, please enter] & The development stage (early concept, prototype, launch, mature product) that shapes research type and scope. \\
        \cmidrule{2-4}
        & Value of Studies &Value of Studies & The reported value of doing user studies. \\
        \cmidrule{2-4}
        & Observation Relevance &Relevance – [name] & The degree to which a research observation is seen as important, valuable, or actionable. \\
    \midrule
    \end{tabular}
    }
 \end{table*}
\clearpage
\begin{table*}[htbp!]
\centering
  \caption{Codebook (continued)}
  \label{apendix:codebook_4}
  \resizebox{\textwidth}{!}{%
  \begin{tabular}{p{2cm} p{2cm} p{3cm} p{5cm}}
    \toprule
    \textbf{Theme} & \textbf{Sub-Theme} & \textbf{Code(s)} & \textbf{Operational Definition} \\
    \midrule
    \multirow{2}{1pt}{Information and Knowledge Acquisition} 
        & Source & Source -[e.g. meetings, clients, developers, SMEs, product documentation, product manager, insights hub, Internet search / Google, people on the team, LLMs, UXRs] & Participant describes sources of information or knowledge such as, meetings, clients, developers, SMEs, product documentation, product manager, insights hub, Internet search / Google, people on the team, LLMs, UXRs etc. \\
        \cmidrule{2-4}
        & Challenges & Challenge - [name] & Mentions of any challenges, pointpoints, obstacles in acquiring information or knowledge. \\
        \cmidrule{2-4}
        & Strategy & [e.g. People-Centric, Document-Centric, etc. whatever fits] & Emphasis on user-centered approaches for acquiring knowledge (e.g., engaging with real users) versus documents and other sources. \\
    \midrule
    \multirow{2}{1pt}{Tool Use} 
        & Tools & Tools – [tool name] & Specific tools used for user research (e.g., Figma, Box, Airtable, Excel, etc.) \\
        \cmidrule{2-4}
        & Usage & Usage – [describe usage] & Observed patterns of tool usage. \\
        \cmidrule{2-4}
        & Future Usage & Usage – [describe usage] & Observed patterns of tool future usage. \\
        \cmidrule{2-4}
        & Tool Challenges & Challenge - [name] & Any difficulty participants face in using tools, e.g.
Lack of Synthesis: Tools failing to support data synthesis effectively. \\
        \cmidrule{2-4}
        & Tool Opportunities & Capabilities - [enter here, e.g. visualization] & Opportunities for using new tools and capabilities (non-AI) e.g. visualization as a tool can improve or expand capabilities. \\
    \midrule
    \end{tabular}
    }
 \end{table*}
\clearpage
\begin{table*}[htbp!]
\centering
  \caption{Codebook (continued)}
  \label{apendix:codebook_5}
  \resizebox{\textwidth}{!}{%
  \begin{tabular}{p{2cm} p{2cm} p{3cm} p{5cm}}
    \toprule
    \textbf{Theme} & \textbf{Sub-Theme} & \textbf{Code(s)} & \textbf{Operational Definition} \\
    \midrule
    \multirow{2}{1pt}{AI Use and Trust Perceptions} 
        & \multirow{2}{1pt}{Current AI Usage} 
            & Usage - [enter here, e.g. Summarization, Domain Knowledge, Planning, Teaching, Discussion Guide, Managing Prompts…] & \textit{Summarization:} Use of AI to condense information.
\textit{Domain Knowledge:} Use of AI to provide or integrate domain-specific knowledge.
\textit{Planning:} Use of AI to support the planning of research or study workflows.
\textit{Teaching:} AI assisting in teaching or training research methods.
\textit{Discussion Guide:} AI helping create or structure discussion guides or protocols.
\textit{Managing Prompts:} Participant strategies managing or sharing prompts. \\   
            \cmidrule{3-4}
            & & Process & Participants have established a specific process or workflow on how to use AI to aid them with user research. \\
            \cmidrule{3-4}
            & & Satisfaction & Participant expressed satisfaction with current use of AI. \\
            \cmidrule{3-4}
            & & Challenges - [name the challenge, concern, or painpoint] & Challenges faced when using or integrating AI into research workflows, including painpoints. \\
        \cmidrule{2-4}
        & \multirow{2}{1pt}{Future AI Usage} 
            & Capabilities – [name] & AI tools and capabilities participants envision adopting in the future. \\    
            \cmidrule{3-4}
            & & Excitement - [Specify If needed] & Anticipated enthusiasm about AI’s future role. \\
            \cmidrule{3-4}
            & & Usage - [Structuring and Grouping, High-Level Overviews, Persona Development, Searchable Repository, …] & \textit{Structuring and Grouping:} Expectation that AI can support clustering and grouping of qualitative data.
\textit{High-Level Overviews:} Anticipation that AI can generate concise summaries or overviews of data.
\textit{Persona Development:} AI envisioned as supporting persona creation.
\textit{Searchable Repository:} AI imagined as enabling efficient search across research knowledge bases. \\
            \cmidrule{3-4}
            & & Challenges - [name the challenge, concern, or painpoint] & Participants expressed concerns or described challenges about future use of AI. \\
        \cmidrule{2-4}
    \end{tabular}
    }
 \end{table*}
\clearpage
\begin{table*}[htbp!]
\centering
  \caption{Codebook (continued)}
  \label{apendix:codebook_6}
  \resizebox{\textwidth}{!}{%
  \begin{tabular}{p{2cm} p{2cm} p{3cm} p{5cm}}
    \toprule
    \textbf{Theme} & \textbf{Sub-Theme} & \textbf{Code(s)} & \textbf{Operational Definition} \\
    \midrule
    \multirow{2}{1pt}{AI Use and Trust Perceptions} 
        & \multirow{2}{1pt}{AI Trust} 
            & Level - [High/Medium/Low] & Participant perceptions of how trustworthy AI is in research contexts. \\    
            \cmidrule{3-4}
            & & Ethical Considerations & Perceptions, concerns, or discussions related to the ethical implications of using AI in the research process. This includes issues such as fairness, bias, transparency, accountability, privacy, and the responsible use of AI-generated insights. \\
            \cmidrule{3-4}
            & & Need for Agency \& Human Control & Desire or need for human oversight, agency, and decision-making in AI-supported processes. \\
        \cmidrule{2-4}
        & Human Value and Experience & Human Value \& Experience &Recognition of unique human contributions and experiential knowledge beyond what AI provides. \\
    \midrule
    \multirow{2}{1pt}{Stakeholder Collaboration} 
        & Goals and Activities & [Guiding, Leading, Sharing, Consulting, Receiving Information, Results Deliberation etc.] & The specific purposes and activities that shape interactions between two or more stakeholders during the research process. This includes roles such as guiding or leading discussions, sharing knowledge or resources, consulting for expertise, receiving updates or information, and collectively deliberating on research findings or decisions. \\
        \cmidrule{2-4}
        & Benefits & [e.g. uncovering unknowns, demonstrating value and craft, ….] & Mention of benefits of working with stakeholders and team members. \\
        \cmidrule{2-4}
        & \multirow{2}{1pt}{Stakeholder Needs and Expectations} 
            & General & Broad expectations or needs expressed by stakeholders without mentioning specific role. \\    
            \cmidrule{3-4}
            & & Need Role - [e.g. Executive, Uplines, UX Researcher, Product Manager, Designer, Senior UX , etc.] & Expectations or needs expressed specifically by a certain role, e.g. executives and upline managers, PMs, UXRs etc. \\
            \cmidrule{3-4}
    \end{tabular}
    }
 \end{table*}
\clearpage
\begin{table*}[htbp!]
\centering
  \caption{Codebook (continued)}
  \label{apendix:codebook_7}
  \resizebox{\textwidth}{!}{%
  \begin{tabular}{p{2cm} p{2cm} p{3cm} p{5cm}}
    \toprule
    \textbf{Theme} & \textbf{Sub-Theme} & \textbf{Code(s)} & \textbf{Operational Definition} \\
    \midrule
    \multirow{2}{1pt}{Stakeholder Collaboration} 
        & \multirow{2}{1pt}{Stakeholder Needs and Expectations} 
            & Need Type - [e.g. Evidence, Delivery Times, Traceability of Insights, Alignment, etc.] & \textit{Evidence:} Emphasis by stakeholders to provide and receive evidence for the outcome of the study. Often video or audio clips of recordings.
\textit{Delivery Times:} Emphasis on fast turnaround of research results.
\textit{Traceability of Insights:} The data presented needs to be connected and traceable, e.g. people want to move from topics to quotes to participants. and so on.
\textit{Alignment:} Alignment of goals and time lines for a user study. \\
        \cmidrule{2-4}
        & \multirow{2}{1pt}{Stakeholders} 
            & 3-in-a-box & Collaboration model involving three primary stakeholders (commonly PM, Design, Research). \\    
            \cmidrule{3-4}
            & & Names - [PM, Design, Research,...] & Stakeholder roles \\
        \cmidrule{2-4}
        & Collaboration Preference & Preference - [e.g. Private Analysis, etc. … enter as you see fit] &Preference for conducting analysis, e.g. individually before sharing with others, or shared with others. \\
        \cmidrule{2-4}
        & Touchpoints & Touchpoint - [e.g. Assumptions \& Questions, Domain Knowledge Transfer, Checkpoint Reviews, Completed Analysis, or others as you see fit] &Specific moments of interaction with stakeholders (e.g., assumptions/questions, domain knowledge transfer, reviews etc.). \\
        \cmidrule{2-4}
        & Collaboration Strategy & Strategy - [e.g. Hands-Off, Involved, or others] &Stakeholder approach to collaboration, e.g.characterized by limited involvement until results are needed, or hands-on, involved, etc. \\
        \cmidrule{2-4}
        & Collaboration Challenges & Challenge - [enter type of challenge, e.g. Time, Resources etc.] &Challenges and frustrations stemming from limited availability, competing priorities, lack of resources, or other challenges. \\
        \cmidrule{2-4}
        & Role Boundaries and Responsibilities & [list any details, e.g. recruiting, Jira tickets, ..] &Clarity (or lack thereof) about stakeholder responsibilities and limits of their involvement. \\
        \cmidrule{2-4}
    \end{tabular}
    }
 \end{table*}
\clearpage
\begin{table*}[htbp!]
\centering
  \caption{Codebook (continued)}
  \label{apendix:codebook_8}
  \resizebox{\textwidth}{!}{%
  \begin{tabular}{p{2cm} p{2cm} p{3cm} p{5cm}}
    \toprule
    \textbf{Theme} & \textbf{Sub-Theme} & \textbf{Code(s)} & \textbf{Operational Definition} \\
    \midrule
    \multirow{2}{1pt}{Stakeholder Collaboration} 
        & Stakeholder Future Opportunities & Opportunity - [e.g. Information Sharing \& Search, etc.] &Name any opportunities for improved stakeholder collaboration, communication, etc. \\
    \midrule
    \multirow{2}{1pt}{UI Feedback Satisfaction} 
        & \multirow{2}{1pt}{Topic representation} 
            & Topic control & Having control to group, edit, move the topics. \\
            \cmidrule{3-4}
            & & Size meaning & Meaning of the size of the bubbles. \\
            \cmidrule{3-4}
            & & Dashed line meaning & Meaning of the color of the bubbles. \\
            \cmidrule{3-4}
            & & Distance meaning  & Participants attributed a meaning to the bubbles distance. \\
            \cmidrule{3-4}
            & & Number of topics in the view & Explanation of how many topics they would be like to see and analyze, in the topic view.  \\
            \cmidrule{3-4}
            & & Topic Relationship & Suggesting how to relate one bubble to another visually. \\
            \cmidrule{3-4}
            & & Challenges & List any other challenges related to the topic representation. \\
        \cmidrule{2-4}
        & \multirow{2}{1pt}{Topic description panel} 
            & Topic Name & Perceptions of the topic name \\
            \cmidrule{3-4}
            & & Topic description & Perceptions of topic description text. \\
            \cmidrule{3-4}
            & & Related Research Objectives & Perceptions of research objectives and value in the analysis.  \\
            \cmidrule{3-4}
            & & Frequency relevance & Perceptions of the numbers in "Frequency". \\
            \cmidrule{3-4}
            & & Importance participant quotes & When participants report they want the participant quotes displayed in the description of the code. \\
            \cmidrule{3-4}
            & & Add transcript & Perceptions and value of adding transcripts. \\
            \cmidrule{3-4}
            & & Find similar quotes & Perceptions and value of finding similar quotes. \\
            \cmidrule{3-4}
            & & Topic coherence understanding & Comprehension of "Topic coherence" term and definition. \\
            \cmidrule{3-4}
            & & Topic strength understanding & Comprehension of "Topic strength" term and definition \\
            \cmidrule{3-4}
            & & Topic strength versus frequency & Mentioned and compared topic strength to frequency. \\
            \cmidrule{3-4}
    \end{tabular}
    }
 \end{table*}
\clearpage
\begin{table*}[htbp!]
\centering
  \caption{Codebook (continued)}
  \label{apendix:codebook_9}
  \resizebox{\textwidth}{!}{%
  \begin{tabular}{p{2cm} p{2cm} p{3cm} p{5cm}}
    \toprule
    \textbf{Theme} & \textbf{Sub-Theme} & \textbf{Code(s)} & \textbf{Operational Definition} \\
    \midrule
    \multirow{2}{1pt}{UI Feedback Satisfaction} 
        & \multirow{2}{1pt}{Topic description panel} 
            & Explanations ratings & When participants wanted explanations for the ratings (low, medium, strong etc.) \\
            \cmidrule{3-4}
            & & Review Strategy & Mentions of how people would review the topic, validate it, what's missing to properly validate, etc. \\
            \cmidrule{3-4}
            & & Add tags purpose & Understanding the value of adding tags. \\
            \cmidrule{3-4}
            & & Metrics suggestions & New suggestions for metrics. \\
            \cmidrule{3-4}
            & & Challenges - [name the challenges] & Participants had difficulties understanding certain elements of the description panel. \\
        \cmidrule{2-4}
        & \multirow{2}{1pt}{Interaction controls} 
            & Approve & Reasons why participants want to approve a topic. \\
            \cmidrule{3-4}
            & & Prefer hide instead of remove topic & Reasons why participants want to hide a topic. \\
            \cmidrule{3-4}
            & & Remove topic & Reasons why participants want to delete a topic. \\
            \cmidrule{3-4}
            & & Expect AI real time update & Expect AI real time update of clusters if they edit/add/interact with topics. \\
            \cmidrule{3-4}
            & & download & Not clear purpose\\
        \cmidrule{2-4}
        & \multirow{2}{1pt}{UI views} 
            & Topic view & Purpose, value, usage of Topic view. \\
            \cmidrule{3-4}
            & & Outliers  value & Purpose, value, usage of outliers. \\
            \cmidrule{3-4}
            & & Automatic outliers & Automatic detection of outliers. \\
            \cmidrule{3-4}
            & & Research objectives & Purpose, value, usage of research objectives. \\
            \cmidrule{3-4}
            & & Integrate views & Want to see outlier view, and/or topic view, and/or research objectives view in the same view.  \\
            \cmidrule{3-4}
            & & Filters Sort Pivot & Would like to filter, sort, or pivot topics, research objectives in the main view. \\
            \cmidrule{3-4}
            & & Visual configuration & When participant reports they want another visual configuration of the displayed element.  \\
            \cmidrule{3-4}
            & & Suggestion new view & When participant suggests a new view, such as a checklist. \\
            \cmidrule{3-4}
    \end{tabular}
    }
 \end{table*}
\clearpage
\begin{table*}[htbp!]
\centering
  \caption{Codebook (continued)}
  \label{apendix:codebook_10}
  \resizebox{\textwidth}{!}{%
  \begin{tabular}{p{2cm} p{2cm} p{3cm} p{5cm}}
    \toprule
    \textbf{Theme} & \textbf{Sub-Theme} & \textbf{Code(s)} & \textbf{Operational Definition} \\
    \midrule
    \multirow{2}{1pt}{UI Feedback Satisfaction} 
        & \multirow{2}{1pt}{UI views} 
            & General & Anything else related to the views we created not listed above. Please name the view. \\
            \cmidrule{3-4}
            & & Challenges - [name the challenges] & Any challenges related to the UI views. \\
        \cmidrule{2-4}
        & \multirow{2}{1pt}{Transcriptions value} 
            & Participant summary & Purpose, value, usage of the participation summary. \\
            \cmidrule{3-4}
            & & Transcript summary & Purpose, value, usage of the transcript summary. \\
            \cmidrule{3-4}
            & & Quotes & Purpose, value, usage of the quotes. \\
            \cmidrule{3-4}
            & & Interaction text transcript & Participant wants to interact (copy, drag \& drop, text from the transcript to other parts of the interface) or want the transcript to be connected to other parts of the UI, e.g. topic description etc. \\
            \cmidrule{3-4}
            & & Transcript visual configuration & When participants report they want to focus on the transcript and not see the other elements on the interface if they don't need those at a specific time.  \\
            \cmidrule{3-4}
            & & Highlight codes Transcript & Want to see codes highlighted in the transcript text. \\
            \cmidrule{3-4}
            & & Suggestions & Participants suggesting new capabilities for the transcript area. \\
            \cmidrule{3-4}
            & & Challenges - [name the challenges] & People encountering any challenges with the transcript area. \\
            \cmidrule{3-4}
        \cmidrule{2-4}
        & \multirow{2}{1pt}{UI chat} 
            & Search info & Use chat as a search for research objectives, keywords, quotes the participant heard during the interviews. \\
            \cmidrule{3-4}
            & & Ask for insights & When participant wants the AI to generate the insights. \\
            \cmidrule{3-4}
            & & Chat usefulness & Participant reports on usefulness and challenges of chat. \\
            \cmidrule{3-4}
            & & Copy chat answer & When participant wants to copy answers or quotes in the chat conversation to topics. \\
            \cmidrule{3-4}
    \end{tabular}
    }
 \end{table*}
\clearpage
\begin{table*}[htbp!]
\centering
  \caption{Codebook (continued)}
  \label{apendix:codebook_11}
  \resizebox{\textwidth}{!}{%
  \begin{tabular}{p{2cm} p{2cm} p{3cm} p{5cm}}
    \toprule
    \textbf{Theme} & \textbf{Sub-Theme} & \textbf{Code(s)} & \textbf{Operational Definition} \\
    \midrule
    \multirow{2}{1pt}{UI Feedback Satisfaction} 
        & \multirow{2}{1pt}{UI chat} 
            & Useful debriefing & The chat would be useful for debriefing discussions. \\
            \cmidrule{3-4}
            & & Prompt guidance & The chat could guide people to write a prompt that will result in expected answers. \\
            \cmidrule{3-4}
            & & Interaction text chat & Participant wants to interact (copy, drag \& drop, text from the chat to other parts of the interface). \\
            \cmidrule{3-4}
            & & Transparency chat & Participant mentions a lack of clarity of the source of information in the chat. \\
            \cmidrule{3-4}
            & & Questions & Participant mentions what kind of questions they would ask in the chat. \\
            \cmidrule{3-4}
            & & Challenges - [name the challenges] & Any other challenges, concerns, difficulties with chat. \\
            \cmidrule{3-4}
            & & Take action & Participant wants the chat to take actions (add tags, remove/hide topics, etc) \\
        \cmidrule{2-4}
        & \multirow{2}{1pt}{UI envisioning usage} 
            & Features envisioned & Suggestions of new features \\
            \cmidrule{3-4}
            & & Integration of features & Participant wants to integrate elements of the UI with one another, e.g. transcript and topics, chat and transcript, etc. \\
            \cmidrule{3-4}
            & & Debriefing  usage & Envisioning the use of the probe in debriefing sessions with the team. \\
            \cmidrule{3-4}
            & & Communication usage & Envisioning the use of the probe in team sessions for communicating the results. \\
            \cmidrule{3-4}
            & & Similar to other tools  & Participant mentions other TA tools with similar features. \\
            \cmidrule{3-4}
            & & Usability - Friendly & Low learning curve to use the probe.  \\
            \cmidrule{3-4}
            & & Usage Goals & Low learning curve to use the probe.  \\
        \cmidrule{2-4}
        & \multirow{2}{1pt}{Collaboration} 
            & Ownership of the analysis & UX should be the one able to modify the analysis into the tool. \\
            \cmidrule{3-4}
            & & Accountability & Who should be accountable to interact with the tool and make changes. \\
            \cmidrule{3-4}
            & & Team collaboration & The tool should be collaborative so that the team can use it jointly and allow sharing of results and insights. \\
            \cmidrule{3-4}
    \end{tabular}
    }
 \end{table*}
\clearpage
\begin{table*}[htbp!]
\centering
  \caption{Codebook (continued)}
  \label{apendix:codebook_12}
  \resizebox{\textwidth}{!}{%
  \begin{tabular}{p{2cm} p{2cm} p{3cm} p{5cm}}
    \toprule
    \textbf{Theme} & \textbf{Sub-Theme} & \textbf{Code(s)} & \textbf{Operational Definition} \\
    \midrule
    \multirow{2}{1pt}{UI Feedback Satisfaction} 
        & \multirow{2}{1pt}{Collaboration} 
            & Access profile & People should have personal profiles to interact with the tool \\
            \cmidrule{3-4}
            & & Receive notifications updates & Would like to get notifications/approvals before any changes are made by collaborators. \\
        \cmidrule{2-4}
        & \multirow{2}{1pt}{AI Expectation in TA} 
            & Productivity & AI should be able to retrieve the context to save time. \\
            \cmidrule{3-4}
            & & Accuracy & AI results should be accurate. \\
            \cmidrule{3-4}
            & & Give insights & AI should give insights in the analysis. \\
            \cmidrule{3-4}
            & & Human Control \& Agency & Humans should be able to own and control the outcome of the AI. \\
            \cmidrule{3-4}
            & & TA guidance process & Ai should help to conduct the analysis in a rigorous way. \\
            \cmidrule{3-4}
            & & Trust & AI should should be safe and show it to the user in the interface. \\
            \cmidrule{3-4}
            & & Remove distractions & AI should remove distractions like things not related to the aim of the analysis. \\
            \cmidrule{3-4}
            & & Flag important content & AI should flag important content. \\
            \cmidrule{3-4}
        \cmidrule{2-4}
        & \multirow{2}{1pt}{UI Overall} 
            & Satisfaction & Participants express overall satisfaction with the UI and experience. This may also include specific features they love. \\
            \cmidrule{3-4}
            & & Dissatisfaction & Participants express concerns or dislikes of the UI or specific features. \\
            \cmidrule{3-4}
            & & Trust & Participants express opinions and thoughts about trust in regard to certain UI features or the overall UI. \\
            \cmidrule{3-4}
            & & Challenges - [name the challenges] & Overall challenges people see with our design. Any challenges that don't fit the other categories. \\
    \bottomrule
    \end{tabular}
    }
 \end{table*}

\section{Details of System validated in Study2}
\label{appendix:study2details}
\subsection{Method}
\begin{algorithm}[H]
\scriptsize
    \caption{Method}\label{alg:system}
    \begin{algorithmic}[1]
        \State \textbf{Input:} Interview transcript, list of Research Objectives (RO), number of topics to be generated per statement (t), conversation context window (c)
        \State \textbf{Output:} Interview transcript with interviewee statements clustered with others of similar meaning.
        \For{n in range(0,length(transcript)}
            \If{speaker is interviewee}
                \State Extract (up to t) topics from the statement, which are higher level theme, concept, or observation; such that each topic is related to the RO. Extract a phrase that supports each topic.
                \State Return a list of the format:
                [\{"topic": "TOPIC1", "phrase": "PHRASE1", "research\_objective": "research objective"\}, \{"topic": "TOPIC2", "phrase": "PHRASE2", "research\_objective": "research objective"\}]
            \EndIf
        \EndFor
        \State Detect clusters of topics using HDBSCAN.
        \State Generate a name and summary for each cluster.
    \end{algorithmic}
\end{algorithm}

\subsection{Prompt for extracting Topics with Llama-70b-instruct}
\label{appendix:prompt_topic_extraction}
\begin{Verbatim}[tabsize=4, breaklines=true]
You are an experienced user researcher and you are very familiar with analyzing interview data. 
An important step in analyzing interview data is to extract TOPICS for each statement made by the interviewee and PHRASES from the statement that support the TOPIC. You will extract TOPICS ONLY for the statement below but as you do this consider the conversation between interviewee and interviewer preceeding the statement.

For this task, strictly follow the INSTRUCTIONS below. 

## CONVERSATION
{conversation}

## INSTRUCTIONS
- Please look at the statement and extract TOPICS. 
- A TOPIC is a higher level theme, concept, or observation. I'm not interested in just keywords. 
- Extract between (0) ZERO and FIVE(5) TOPICS but no more than that. Pick what's most important for the statement.
- The TOPICS of the statement need to be related or in support of the RESEARCH OBJECTIVES. If they are not related do not force topic extraction.
- Extract a long PHRASE from the statement that best supports each TOPIC.
- For each TOPIC, also select ONE of the below RESEARCH OBJECTIVES it is most related to.
- Extract TOPICS ONLY for the statement, but as you do this, consider the CONVERSATION between interviewee and interviewer preceeding the statement as additional context.
- Do not generalize.
- Do not use the research objectives literally as topics
- Do not include any introductory text or explanations in your RESPONSE — your output is only the list of TOPICS and their associated PHRASES as described under OUTPUT FORMAT and an empty JSON object if no match is found.


## RESEARCH OBJECTIVES
{research_objectives}

## OUTPUT FORMAT
Return a valid json object containing the topics and their phrases, e.g.:
[
  {"topic": "TOPIC1", "phrase": "PHRASE1", "research_objective": "research objective id (if available) and full name"},
  {"topic": "TOPIC2", "phrase": "PHRASE2", "research_objective": "research objective id (if available) and full name"},
  {"topic": "TOPIC3", "phrase": "PHRASE3", "research_objective": "research objective id (if available) and full name"},
  {"topic": "TOPIC4", "phrase": "PHRASE4", "research_objective": "research objective id (if available) and full name"},
  {"topic": "TOPIC5", "phrase": "PHRASE5", "research_objective": "research objective id (if available) and full name"}
]

If no match of a TOPIC with any RESEARCH OBJECTIVES is found, return:
[]
\end{Verbatim}

\subsection{Prompt for using an LLM to extract similar and unique items from two lists}
\label{appendix:prompt_topic_list_compare}
\begin{Verbatim}[tabsize=4, breaklines=true]
<|begin_of_text|><|start_header_id|>user<|end_header_id|>
Compare LIST_A and LIST_B. 
Identify the items which are present in both lists and which are unique to each list. 
When trying to find items that are part of both lists, you may not find exact matches, hence look for semantic similarity between items.
When two items are found to be similar, return them as a sub-list.
If any item from LIST_A or LIST_B is found to be similar to an item from the other list, it cannot be included in the unique items list again.
For this task, strictly follow the OUTPUT FORMAT below. Do not return any text apart from the output format.

## LIST_A
{list_A}

## LIST_B
{list_B}

## OUTPUT FORMAT
Return a valid json object containing the topics and their phrases, e.g.:
[
    {
        "present_in_both_lists": [
            ["ITEM1", "ITEM4"]
        ],
        "unique_items_in_list_A: [
            "ITEM3"
        ],
        "unique_items_in_list_B: [
            "ITEM2", "ITEM5"
        ]
    }
]

RESPONSE:
<|eot_id|><|start_header_id|>assistant<|end_header_id|>
\end{Verbatim}

\subsubsection{Prompt for generating a name representative of the topics in the clusters}
\label{appendix:prompt_topic_cluster_name}
\begin{Verbatim}[tabsize=4, breaklines=true]
<|begin_of_text|><|start_header_id|>user<|end_header_id|>
You are an experienced user researcher and you are very familiar with analyzing interview data. An important step in analyzing interview data is to label the group of items in the TOPICS cluster list. To complete the task, use TOPICS and follow the INSTRUCTIONS below. 

## TOPICS
{chunk}

## INSTRUCTIONS
- Please look at the TOPICS and assign a short label for all the items in the TOPICS list

## OUTPUT FORMAT
Return a valid string that represents the short name of the TOPICS cluster. Do not provide an explanation or any other oputput. Only the name of the topic cluster.

If no suitable label is found, return: ""

RESPONSE:
<|eot_id|><|start_header_id|>assistant<|end_header_id|>
\end{Verbatim}

\subsubsection{Prompt for generating a summary representative of the topics in the clusters}
\label{appendix:prompt_topic_cluster_summary}
\begin{Verbatim}[tabsize=4, breaklines=true]
<|begin_of_text|><|start_header_id|>user<|end_header_id|>
You are an experienced user researcher and you are very familiar with analyzing interview data. An important step in analyzing interview data is to describe the group of items in the TOPICS cluster list. To complete the task, use TOPICS and follow the INSTRUCTIONS below. 

## TOPICS
{chunk}

## INSTRUCTIONS
- Please look at the TOPICS and assign a short summary for all the items in the TOPICS list

## OUTPUT FORMAT
Return a valid string that represents the short summary of the TOPICS cluster. Do not provide an explanation or any other output. Only the summary of the topic cluster.

If no suitable label is found, return: ""

RESPONSE:
<|eot_id|><|start_header_id|>assistant<|end_header_id|>
\end{Verbatim}

\end{document}